\documentclass[12pt,preprint]{aastex}

\usepackage{amstex}

\newcommand{\hk}{\mathrm{H}-\mathrm{K}}
\newcommand{\ehk}{\mathrm{E}(\mathrm{H}-\mathrm{K})}

\shorttitle{Structure of B335}
\shortauthors{Harvey et al.}

\begin{document}

%\slugcomment{Draft: \today}

\title{Structure of Protostellar Collapse Candidate B335\\ Derived
from Near-Infrared Extinction Maps \footnote{Based on observations
with the NASA/ESA Hubble Space Telescope, obtained at the Space
Telescope Science Institute, which is operated by the Association of
Universities for Research in Astronomy, Inc.  under NASA contract No.\
NAS5-26555, and observations obtained at the W.M.\ Keck Observatory,
which is operated as a scientific partnership among the California
Institute of Technology, the University of California and NASA, made
possible by the generous financial support of the W.M.\ Keck
Foundation.}}

\author{Daniel W.A.\ Harvey\altaffilmark{2}, 
	David J.\ Wilner\altaffilmark{3}, 
	Charles J.\ Lada\altaffilmark{4},
	Philip C.\ Myers\altaffilmark{5}}

\affil{Harvard-Smithsonian Center for Astrophysics, 60 Garden Street,
Cambridge, MA 02138}

\author{Jo\~{a}o F.\ Alves\altaffilmark{6}}

\affil{European Southern Observatory, Karl-Schwarzschild Stra{\ss}e 2,
D-85748 Garching bei M\"{u}nchen, Germany}

\and

\author{Hua Chen\altaffilmark{7,8}}

\affil{Steward Observatory, University of Arizona, Tucson, AZ 85721}

\altaffiltext{2}{dharvey@@cfa.harvard.edu}
\altaffiltext{3}{dwilner@@cfa.harvard.edu}
\altaffiltext{4}{clada@@cfa.harvard.edu}
\altaffiltext{5}{pmyers@@cfa.harvard.edu}
\altaffiltext{6}{jalves@@eso.org}
\altaffiltext{7}{huachen@@nortelnetworks.com}
\altaffiltext{8}{Current address: Nortel Networks, 2305 Mission
College Blvd, Santa Clara, CA 95954}

\begin{abstract}
We present a near-infrared extinction study of the dark globule B335,
a protostellar collapse candidate, using data from HST/NICMOS and the
W.M.\ Keck Observatory. These data allow a new quantitative test of
the ``inside-out'' collapse model previously proposed to explain
molecular line profiles observed toward this region. We find that the
{\em shape} of the density profile is well matched by the collapse
model, but that the {\em amount} of extinction corresponds to larger
column densities than predicted. An unstable Bonnor-Ebert sphere with
dimensionless outer radius $\xi_{\mathrm{max}}=12.5\pm2.6$ provides
an equally good description of the density profile, and is 
indistinguishable from the collapse model over the range in radius
sampled by the extinction data. The bipolar outflow driven by the
embedded young stellar object has an important effect on
the extinction through the core, and modeling the outflow as a
hollowed-out bipolar cone of constant opening angle provides a good
match to the observations. The complete exinction map is well
reproduced by a model that includes both infall and outflow, and an
additional 20\% dispersion that likely results from residual turbulent
motions. This fitted model has an infall radius of 
$R_{\mathrm{inf}} = 26 \pm 3''$
($0.031$~pc for 250~pc distance), and an outflow cone semi-opening angle 
of $\alpha = 41 \pm 2^{\circ}$. The fitted infall radius is consistent
with those derived from molecular line observations and supports the
inside-out collapse interpretation of the density structure. 
The fitted opening angle for the outflow is slightly larger than observed 
in high velocity CO emission, perhaps because the full extent of the 
outflow cone in CO becomes confused with ambient core emission at low
velocities.
\end{abstract}


\section{Introduction}
The near-infrared color excess of extincted background stars can
reveal the dust distribution in a dense core and provide information
on physical structure. Historically, this technique has been hampered
in small, dense regions by limited instrumental sensitivity (Jones et
al.\ 1980, 1984), but the development of large format array cameras
has sparked renewed interest (Lada et al.\ 1994, Alves et al.\ 1998,
Lada, Alves \& Lada 1999, Alves, Lada \& Lada 1999, 2001). The
unprecedented sensitivity available with the Near Infrared Camera and
Multi Object Spectrometer (NICMOS) instrument on the Hubble Space
Telescope (HST) opens up a new regime. In particular, the high
extinctions of dense cores that form the immediate environs of
candidate low mass protostars are now accessible.

Molecular line surveys of nearby dark clouds have identified a large
number of low mass dense cores (Myers, Linke \& Benson 1983) of which
roughly half are associated with young stellar objects detected by
{\em IRAS} (Beichman et al.\ 1986). These observations have
contributed to the general paradigm that at least some low-mass stars
like the Sun form by the gravitational collapse of isolated, dense
cores. In the standard theory, originating with Shu (1977), the
dynamical collapse to a star/disk system occurs from the inside-out.
First, an $n\propto r^{-2}$ distribution is established prior to
collapse as the cloud core loses magnetic and turbulent support
through ambipolar diffusion and relaxes to a balance between gravity
and thermal pressure. Collapse is initiated at the center where the
density is highest, and a wave of infall propagates outward at the
sound speed. Conditions inside the infall radius asymptotically
approach free-fall, with $n\propto r^{-1.5}$ and $v\propto r^{-0.5}$.
Modifications to the simple theory have been made to account for more
realistic initial conditions, including slow rotation (Tereby, Shu \&
Cassen 1984) and residual magnetic fields (Galli \& Shu 1993, Li \&
Shu 1997). A similar radial density profile is also produced by the
competing self-similar collapse theory of Larson (1969) and Penston
(1969). However, the Larson-Penston flow predicts far larger infall
velocities than the Shu theory, and has proved unsuccessful at
explaining molecular line observations (see e.g.\ Zhou 1992).

Unambigous identification of inside-out collapse motions has proved
difficult for several reasons. First, the fact that collapse initiates
in the center of a core means that it can be detected only with
observations of sufficiently high angular resolution. Further
difficulty comes from the fact that the velocities associated with
infall motions are small, comparable to intrinsic line widths and much
smaller than the velocities of the bipolar outflows powered by young
stellar objects. The presence of localized, redshifted self-absorption
in spectral lines of moderate optical depth provides one signature of
infall motions, but this signature is not unique. Scenarios other than
collapse can produce the same spectral line profiles. The most
promising claims for collapse have been inferred from redshifted
self-absorption, and consequently these claims are often controversial
(see the review of Myers, Evans \& Ohashi 2000).

The dense core in the B335 region is generally recognized as the best
protostellar collapse candidate. The B335 region contains an isolated,
roughly spherical globule at a distance of roughly 250~pc (Tomita,
Saito \& Ohtani 1979). Deeply embedded in the globule is a young
stellar object discovered at far-infrared wavelengths by Keene et al.\
(1983) and detected only at $\lambda > 60~\mu$m by {\em IRAS}.
Observations of high velocity molecular line emission have shown that
the young stellar object drives a bipolar outflow with semi-opening angle
of $\sim 25\pm5^{\circ}$ (Cabrit, Goldsmith \& Snell 1988, Hirano et
al.\ 1988, 1992). Detailed radiative transfer models based on the theory
of inside-out collapse provide good fits to many molecular line
profiles of dense gas tracers observed at $10''$--$30''$ resolution
(Zhou et al.\ 1993, Choi et al.\ 1995). These models imply an infall
radius of 0.03~pc ($25''$), an age of $1.3 \times 10^{5}$~yrs, and a
central mass of $0.37$~M$_{\odot}$. While the agreement of the model
with few parameters and the observed molecular line profiles strongly
support inside-out collapse, the model is not unique. For example, in
one extreme scenario discussed by Zhou et al.\ (1993), an unrelated
foreground cloud of just the right excitation properties absorbs
emission from a static core and exactly reproduces the infall
signatures. Recent molecular line observations of B335 made with the
higher angular resolution of interferometry are also starting to show
perhaps important discrepancies with the predictions of inside-out
collapse (Wilner et al.\ 2000).

An extinction map of sufficient depth can probe the density structure
of B335 and test the claim for inside-out collapse in a way that is
free from many of the problems that plague molecular line studies,
including excitation effects and spatially varying abundances of the
molecular tracers. B335 is fortuitously placed for an extinction
study, located at Galactic coordinates $l = 44.9$, $b = - 6.6$, which
is a direction with a very high density of background stars with
uniform properties. Previous optical extinction studies of B335 were
not sensitive to the structure of the dense core anywhere close to
where the molecular line data suggest collapse (see Bok \& McCarthy
1974, Dickman 1978, Tomita et al.\ 1979 and Frerking et al.\ 1987). In
this paper, we present a near-infrared extinction study of B335 using
data from HST/NICMOS, supplemented by data from the W.M. Keck
Observatory.  Section~\ref{sec:obs} briefly reviews the technique and
describes the observations and data reduction.
Section~\ref{sec:results} presents an analysis of the extinction data
in the context of several physical models for the B335 density
structure, and comparisons with the results of previous studies.
Simulations of the observations are also presented, using a Monte Carlo
approach detailed in the appendix. Section~\ref{sec:conc} summarizes
the main conclusions.

\section{Observations and Data Reduction}
\label{sec:obs}
\subsection{Review of the Near-Infrared Color Excess Technique}
\label{sec:method}
Measurements of the extinction of stars background to a dense core 
provides estimates of column density along many pencil beams. These 
estimates can be used to determine the overall structure of the
obscuring core. The extinction measurements are best done at infrared
wavelengths, where absorption due to dust is much less than in the
optical, thereby probing regions of higher (visual) extinction. The
basic method is to measure the near-infrared color excess for each
background star:
\begin{equation}
\ehk = (\hk)_{\mathrm{observed}} - (\hk)_{\ast} \ ,
\end{equation}
where $(\hk)_{\ast}$ is the intrinsic color of the star. The color
excess represents the differential optical depth between the two
near-infrared filters, which is directly proportional to dust column
density. The observed color excess may be converted to an equivalent
visual extinction using the standard reddening law (e.g.\ Rieke \&
Lebosfsky 1985), and then converted to gas column density by adopting
a gas-to-dust ratio (e.g.\ Bohlin et al.\ 1978). There are some
uncertainties in both of these conversions. The gas-to-dust ratio in
protostellar cores is the subject of some debate; at low extinction
there is some evidence to suggest that it is constant along different
lines of sight in the galaxy (Bohlin et al.\ 1978), but at high
extinction it is unclear whether the gas-to-dust ratio remains
constant. For example, if grain growth occurs in the high extinction
regions as a result of molecular depletion and agglomeration, then
one might expect lower extinction for a given column density. The
reddening law is probably less problematic and nearly universal at
near-infrared wavelengths, given the small grain sizes present in
protostellar envelopes (Mathis 1990).

In practice, the color excess method does not require a knowledge of
the intrinsic colors of the individual extincted stars. A statistical
correction to their observed colors may be obtained empirically from
the background stellar population, provided it is sufficiently
spatially uniform. For stars of spectral types A0 to late M, in the
main sequence or the giant loci, the intrinsic colors lie in the
narrow range $0 < (\hk)_{\ast} < 0.3$ (Koornneef 1983). Therefore,
at high extinctions, little error is introduced by simply assuming a
mean background star color derived from nearby control fields.

\subsection{NICMOS Observations}
The B335 globule was imaged with the NIC3 camera on HST (Cycle 
7---NICMOS, GO-7843, 1998 June 23) using two broadband filters, F160W
and F222M, similar to the usual H band and K band filters that are
matched to the 1.6~$\mu$m and 2.2~$\mu$m atmospheric windows. The NIC3
camera has $256\times256$ pixels with size $0\farcs20$ giving a
$51\farcs2$ field of view. The observations were obtained during the
second of two periods of a few weeks when the HST secondary mirror was
adjusted to allow the NIC3 camera to be in focus (the ``NIC3
Campaigns''). 

\begin{figure}[htb]
\figurenum{1}
\setlength{\unitlength}{1in}
\begin{picture}(6,6.0)
\put(-0.2,-1.5){\includegraphics{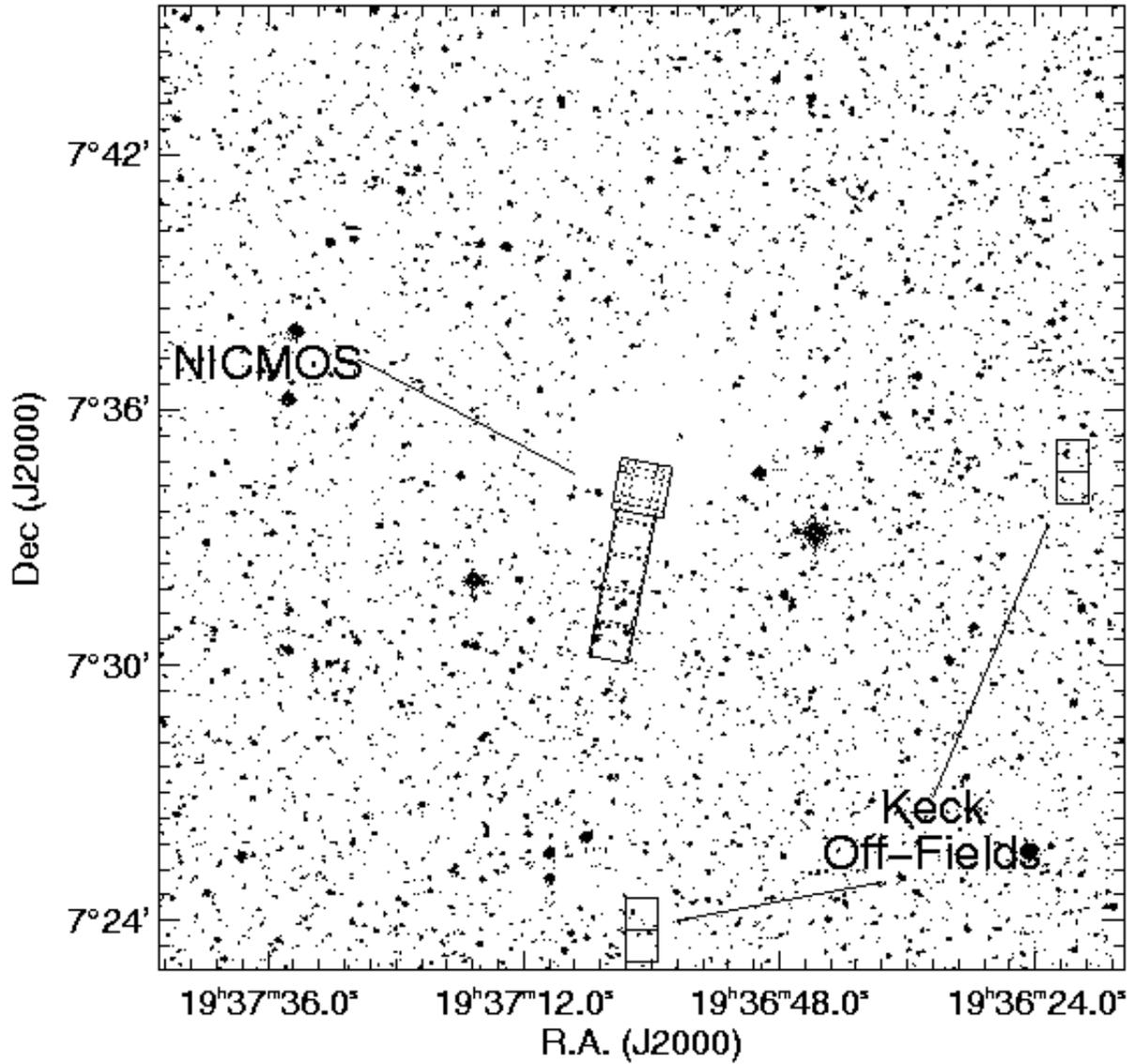}}
\end{picture}
\caption{Wide field optical view of the B335 region from the Digital 
Sky Survey. Superimposed is a diagram of the observed fields from 
both NICMOS and Keck.\label{fig:skysurv}}
\end{figure}

The observing program had two parts: (1) a deep $3\times3$ mosaic of
the central region of high extinction, taking 6 orbits, and (2) a
radial strip of images extending about $4'$ from the center, taking 1
orbit, to reach the background stellar population and provide a sample
of stars to transform NICMOS magnitudes to a standard photometric
system. The length of the strip was the maximum allowed before
suffering the heavy overhead of guide star reaquisition. (The strip
included the center of B335 as insurance in the event that the deeper
observations were unsuccessful.) Figure~\ref{fig:skysurv} shows the
observed fields overlayed on a wide field optical view of the region
from the Digital Sky Survey. We note that the NIC3 orientation angle
on the sky and stepping direction of the radial strip were given by
the spacecraft orientation on the day of observation. Advance
specification of a particular orientation would have created an
untenable scheduling constraint; the orientation obtained was based
solely on where the spacecraft happened to be during the period when
the NIC3 camera was in focus and B335 was accessible.

The central region was observed with a standard nine-point spiral
dither pattern with $10''$ offsets, and a total exposure time of 896
seconds per dither. The resulting square mosaic covers $72''$. The
strip images were taken in {\em one-chop} mode, an exposure time of 
56 seconds per chop, with an offset of $3''$ between chops, and
roughly $3\farcs7$ overlap between successive images. This resulted
in a strip of coverage $55''$ wide that extends $247''$ from the
mosaic center. The data were taken in the NICMOS {\em multiaccum}
mode, for which an initial readout of the detector array is followed
by non-destructive readouts during the course of a single integration.
The intermediate readouts can be used to identify and to remove the
effects of cosmic ray hits and saturated pixels.

The NICMOS data were obtained from the HST archive in FITS format
both unprocessed and already processed through the STScI pipeline
(CALNIC A \& B). Initial inspection showed the STScI pipeline did not
adequately deal with a variety of instrumental effects,  and so we
reprocessed the data in IRAF using the NICREDUCE software developed
by Brian McLeod (McLeod 1997, Lehar et al.\ 1999). The NICREDUCE
package provides superior and more versatile handling of sky and dark
current subtraction, as well as cosmic ray hits and other defects.

The bottom rows of the NICMOS images suffered from a small amount of
vignetting. The effect was most pronounced in the F222M images, where
about $2\farcs4$ were unusable. The overlap of images in the dither
pattern and in successive images of the strip was sufficient to
ensure that the vignetting resulted in only a small loss in
sensitivity in the affected regions, and no gaps in the spatial
coverage.

The dithered images of the central region of B335 suffered an
additional complication due to the presence of a very bright star (H
magnitude of 9.0) on the west side of the mosaic center. The enormous
count rate from the star caused the image reduction software to produce
a small depression in the continuum level of the quadrant of each image
that contained the star. This deviation was fixed by adding a constant to
the relevant quadrant so that for each image the median pixel value
was uniform in all four. Ghost impressions of the star in successive
exposures were interactively masked in the affected images and were
not included in the final mosaic.

\subsubsection{Photometry}
Stars were identified by visual inspection of the images, as several
attempts at automatic star finding were compromised by the effects of
diffraction spikes and rings around the brighter stars. Photometry on
each star was performed with the suite of routines in the {\em apphot}
package. A series of apertures from $0\farcs2$ to $0\farcs5$ in radius
was used for each star. Aperture corrections were then performed using
a minimum of four isolated and bright stars from the relevant image.
Each aperture was used to calculate a magnitude and associated error
as a check to identify any unusual behavior. The measured count rates
were corrected to a $1''$ radius aperture, and then multiplied by
1.075 as detailed in the NICMOS Photometric Calibration CookBook
(available from STScI). The instrumental magnitudes were then
transformed to the Vega scale using constants for each filter provided
in the HST Data Handbook (Version 4, Dec 1999). For stars that were
observed in a region of overlap between strip images, weighted mean
magnitudes were calculated. For stars observed in the overlap region
of the central mosaic and the first strip image, the value from the
more sensitive central mosaic was taken. The complete NICMOS
photometry list contains 1026 stars detected with the F160W filter and
471 stars detected with the F222M filter with magnitude error $<
0.1$. The NICMOS limiting magnitudes are spatially dependent and are
depicted in Figure~\ref{fig:limplot}. In the central 32$''$, with 
the most integration time, the limiting magnitudes are H=24.8 and 
K=21.5 (in the CIT system).

\subsection{Keck Observations}
Observations of B335 were made on 1998 July 10 using NIRC (Matthews \&
Soifer 1994) on the Keck I telescope. These observations were made for
two reasons: (1) to obtain a better study of the stellar background
population around B335, and (2) to transform the NICMOS F160W and
F222M magnitudes to a standard system. The NIRC has
$256\times256$ pixels with size $0\farcs15$, giving a $38\farcs4$
field of view. Each basic observation consisted of nine dithered 5
second integrations, in a $3\times3$ pattern with $3''$ offsets, which
were mosaiced together. Observations were made with J, H, and K
filters. The program had three parts: (1) nine images of the central
region of B335, approximately mimicking the sub-images of the NICMOS
central mosaic, (2) five images leading away from the center of B335,
at each position of a NICMOS strip image, and (3) four additional
fields $10'$ away from B335, to study the characteristics of the
stellar background population.  These regions are marked on
Figure~\ref{fig:skysurv}.

The data reduction and calibration was done in IRAF following the
standard dark/flat/sky subtraction procedures, using a running flat.
This procedure works well when the background emission does not change
rapidly. For some periods during the observations, the sky emission in
the H band did change relatively quickly, and the background
subtraction for observations during these periods is not perfect. The
typical seeing during the night was $0\farcs5$, though there were
periods when it was better and also much worse.

\subsubsection{Photometry}
The majority of stars in the Keck images of B335 could be identified
with reference to the NICMOS data, since the Keck images are all less
deep than the NICMOS equivalents and the field of view slightly
smaller.  For the Keck fields not observed with NICMOS, all stars were
again identified by visual inspection.

Photometry was performed for all non-saturated stars using the {\em
apphot} package. Again, a series of increasing radial apertures was
used for each star, and each aperture was used to calculate a
magnitude.  This approach allowed a confidence level in the
measurement to be assigned to each star, based on whether the
distribution of measurements from different apertures was consistent
with the predicted uncertainty.  If the spread of the various
magnitudes was greater than 0.2 magnitudes, then the measurement was
identified as low quality (18\% of the stars detected at H and 17\% of
the stars detected at K). These low quality stars have
point-spread-functions that differ from the norm, mainly from those
periods of poor and varying seeing conditions. The instrumental
magnitudes were transformed to the CIT system through observations of
the HST infrared standards SJ9164 and SJ9182 (Persson et al.\ 1998)
and the UKIRT standard FS35 (Casali \& Hawarden 1992). The internal
dispersion in the zero points measured for these stars was 0.02
magnitudes, and no atmospheric extinction correction could be
discerned.  The Keck photometry list contains a total of 297 stars
detected at H, and 340 stars detected at K, in the overlap region
between the Keck and NICMOS datasets.

The four Keck fields $10'$ away from B335 contain typically 100
detected stars per square arcminute with H$<20$. The four fields
display uniform properties and no discernable reddening; the star
counts in each image are identical to within Poisson errors, and the
mean $(\hk)$ color of the fields are all in the range $0.07$ to
$0.11$, consistent with the intrinsic colors of giant and main
sequence stars. A detailed study of the background stellar population
is presented in Section~\ref{sec:background}. The limiting magnitudes
of the Keck observations are approximately H=19.7 and K=19.6. 

\subsection{Transformation of NICMOS Magnitudes}
\begin{figure}[htb]
\figurenum{3}
\setlength{\unitlength}{1in}
\begin{picture}(6,6.0)
\put(-0.1,-2.3){\includegraphics{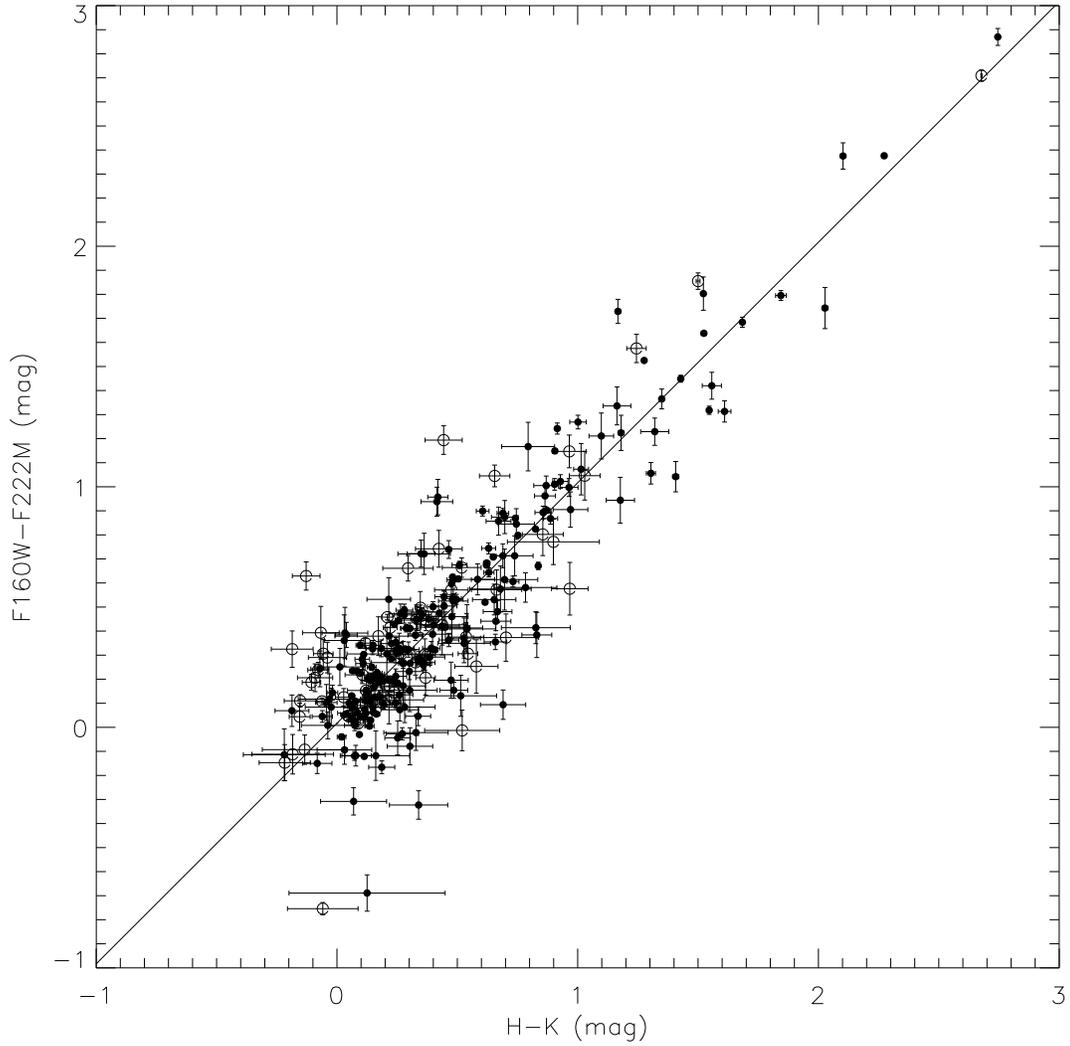}}
\end{picture}
\caption{Plot of NICMOS (F160W${}-{}$F222M) color against Keck 
(H${}-{}$K) color. Stars for which the Keck H or K measurement are 
of lower quality are marked as open circles. Also plotted is the 
derived transformation: $(\mathrm{F160W}-\mathrm{F222M}) = (\hk) + 0.017$.
\label{fig:coltrans}}
\end{figure}

For extinction calculations, the NICMOS F160W and F222M magnitudes are
transformed to standard H and K magnitudes. Linear least squares fits
(not including the lower quality Keck data) give slopes of
$-0.05$ for (H${}-{}$F160W) vs.\ F160W,
$-0.03$ for (K${}-{}$F222M) vs.\ F222M, and
$0.99$ for (H${}-{}$K) vs.\ (F160W${}-{}$F222M), where the formal 
uncertainties in the fitted slopes are $\sim 0.01$. The formal
uncertainties are underestimates of the true uncertainties since the
magnitude comparison slopes are strongly biased by the few brightest
stars, and the color comparison slope is biased by the few reddest
stars. Experimentation with subsets of the stars resulted in
variations in the fitted slopes at the $0.05$ level. We conclude that
these data provide no evidence for non-zero slopes or color terms in
the transformation. We calculated constant offsets for the magnitude
conversions for each filter, using a weighted mean of (H${}-{}$F160W)
and (K${}-{}$F222M) values. 
The result is H = F160W${}+ (0.035 \pm 0.003)$, and
              K = F222M${}+ (0.052\pm 0.003)$, 
where the uncertainties represent the standard error in the mean.
Figure~\ref{fig:coltrans} shows the scatter plot of $(\hk)$ vs.\ 
(F160W${}-{}$F222M) with the derived color transformation
superimposed. For the $(\hk)$ colors of interest, in the range 0 to 
3, the differences between the two photometric systems are remarkably
small.

Support for the accuracy of these transformations comes from published
ground-based photometry of the one very bright star observed by
Kr\"{u}gel et al.\ (1983), who measured Johnson H and K fluxes of
$310\pm25$~mJy and $605\pm45$~mJy, which correspond to magnitudes of
H${}=8.89\pm0.07$ and K${}=7.58\pm0.07$, and color 
$(\hk)=1.31\pm0.11$. These values compare very favorably with the
transformed NICMOS magnitudes of H${}=9.039\pm0.001$ and 
K${}=7.675\pm0.001$, and color $(\hk)=1.364\pm0.001$. 
Note that systematic errors limit the accuracy of the NIC3 photometry to 
a few percent, and are not reflected in the random uncertainties quoted
here.

\section{Results and Analysis}
\label{sec:results}
\subsection{The Background Population Mean Color and Dispersion}
\label{sec:background}
\begin{figure}[htb]
\figurenum{4}
\setlength{\unitlength}{1in}
\begin{picture}(6,4.0)
\put(0.0,-4.3){\includegraphics{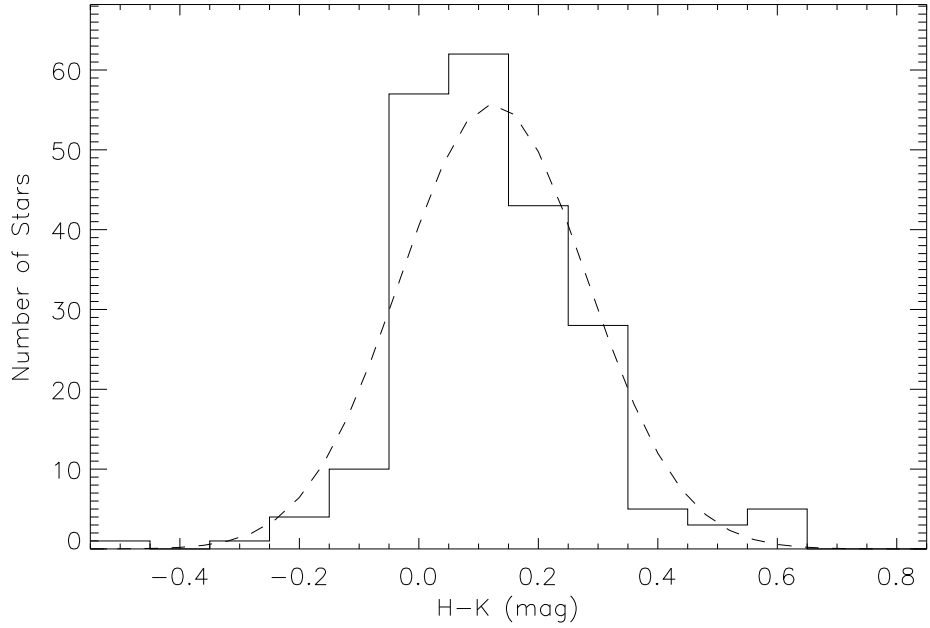}}
\end{picture}
\caption{Histogram of the $(\hk)$ colors of the background 
population, with bins of width 0.1 magnitudes. Also plotted is a
Gaussian distribution with the same mean ($\overline{\hk}=0.13$) and
standard deviation ($\sigma(\hk)=0.16$), and normalized so that the
area under the curves are equal. Stars observed at Keck with lower
quality photometry as discussed in the text are not included in this
histogram.\label{fig:backcolor}}
\end{figure}

To check the uniformity of the background stellar population, the
starcounts and mean colors of the four Keck off fields were compared
with those for the two NICMOS strip images furthest away from the
center of B335. The colors in the different images are consistent
within the observed dispersions. The fifth NICMOS strip field has a
mean color and standard deviation of $\overline{\hk}=0.17$,
$\sigma(\hk)=0.18$, very similar to that found for the combined Keck
off-fields: $\overline{\hk}=0.10$, $\sigma(\hk)=0.13$. The fourth
NICMOS strip field shows a slightly redder mean color
($\overline{\hk}=0.26$) and higher dispersion ($\sigma(\hk)=0.23$)
than the other fields; most likely, this field is not far enough away
from the center of B335 to be completely free of its extinction, and
this field was excluded from our characterization of the background
stellar population. Figure~\ref{fig:backcolor} presents a histogram of
the $(\hk)$ colors of the ``background'' stars in 0.1 magnitude bins
(excluding the Keck stars with lower quality photometry). The $(\hk)$
mean color is 0.13 magnitudes, with standard deviation 0.16
magnitudes.  Also shown for comparison is a Gaussian distribution with
the same mean and standard deviation (normalized so that the area
under each curve is equal). A color-magnitude diagram constructed for
this sample shows no sign of any important systematic trends,
e.g. there is no tendency for fainter stars to be redder. The
starcounts in the fifth NICMOS strip field are indistinguishable from
those in the Keck off-fields, taking into account the different
different size fields of view.

\subsection{NICMOS Central Mosaic Image}
\begin{figure}[htb]
\figurenum{5}
\setlength{\unitlength}{1in}
\begin{picture}(6,4.0)
\put(-0.6,-4.){\includegraphics{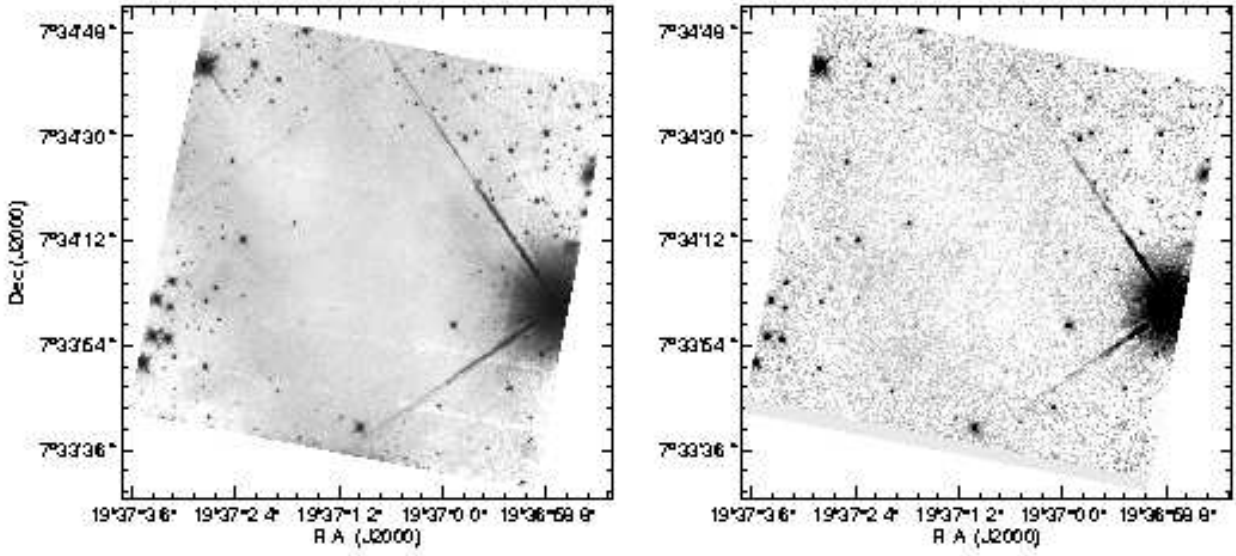}}
\end{picture}
\caption{NICMOS F160W (left) and F222M (right) mosaic images of
the central B335 region.\label{fig:images}}
\end{figure}

\begin{figure}[htb]
\figurenum{6}
\setlength{\unitlength}{1in}
\begin{picture}(6,7.0)
\put(-0.5,-0.3){\includegraphics{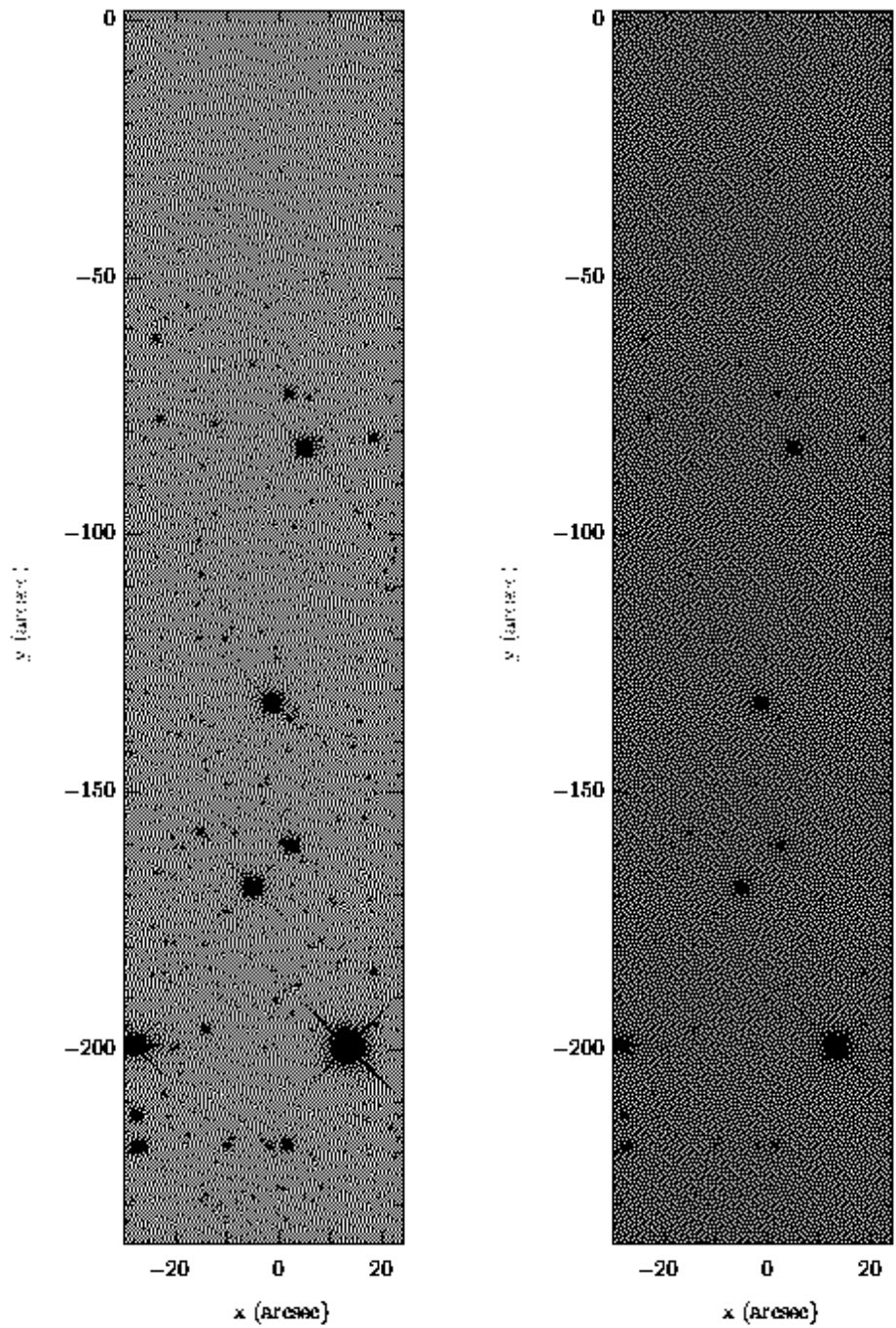}}
\end{picture}
\caption{NICMOS F160W (left) and F222M (right) images of the 
strip leading off B335.\label{fig:strip}}
\end{figure}

Figures~\ref{fig:images}~\&~\ref{fig:strip} show the NICMOS F160W and
F222M images of the B335 region. The central mosaic images in
Figure~\ref{fig:images} may be compared to H band and K band images of
a smaller part of the central region of B335 obtained over three
nights at Keck by Hodapp (1998) using NIRC. In the region of overlap,
the F222M image shows the same stars as the K band image, but the
overall sensitivity of the NICMOS image is better and the
point-spread-function much sharper, leading to better separation of
nearby stars. The F160W image shows many more stars than the
corresponding H band image. In contrast with the ground-based data,
essentially all stars detected with the F222M filter are also detected
with the F160W filter because of the low background and greater
sensitivity of the space-borne NICMOS at the shorter wavelength.

\begin{figure}[h]
\figurenum{7}
\setlength{\unitlength}{1in}
\begin{picture}(6,4.0)
\put(-1.2,-6.){\includegraphics{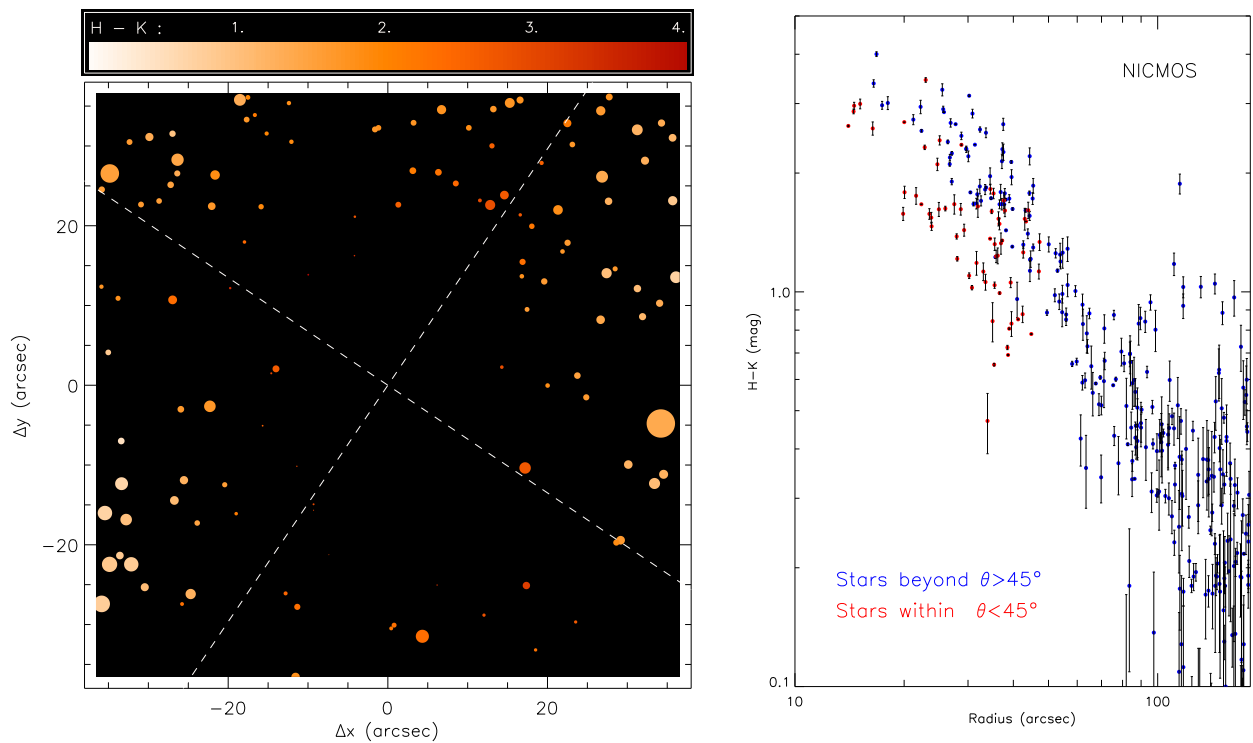}}
\end{picture}
\caption{Left: Pseudo-image of the NICMOS mosaic: the axes are
spatial, magnitude determines a star's ``size'', and $(\hk)$
determines its color. The axes are in the frame of the NICMOS image.
The mosaic contains 119 stars detected in both filters. The dashed
line distinguishes the regions that are more than $45^{\circ}$ from
the outflow axis (the outflow axis has position angle roughly
$100^{\circ}$ in the pseudo-image). Right: $(\hk)$ color against
radius out to $180''$ from the center of B335. There is a steep
gradient towards the protostar location. The symbol color indicates
whether the star lies within $45^{\circ}$ of the outflow axis.
The outer edge of B335 occurs at a radius of approximately
$125''$.\label{fig:nicmos}}
\end{figure}

Figure~\ref{fig:nicmos} conveys much of the information from the central
mosaic of NICMOS observations. The left plot is a pseudo-image in that the
two axes are spatial, the apparent brightness of a star determines its
``size'', and the value of its $(\hk)$ color determines its ``color''. The
right-hand plot displays the radial dependence of the $(\hk)$ colors out
to $180''$ from the center. The origin in both plots is given by the
position of the embedded young stellar object, as determined by
interferometric observations of millimeter continuum emission (Wilner et
al.\ 2000); the uncertainty in the registration of the NICMOS images and
the millimeter peak is estimated to be about $1''$. Based on an inspection
of the extinctions, the stars in the right hand plot have been marked
according to their spatial location: those that lie in the North or South
quadrants are marked in red, and those that lie in the East or West
quadrants are marked in blue (the quadrant boundaries are marked in the
pseudo-image as dashed lines). Defining $\theta$ to be the angle subtended
between the vector to a star and the axis of the outflow (which runs very
close to east-west), this division may be expressed simply as
$\theta > 45^{\circ}$ or $\theta < 45^{\circ}$. The NICMOS
central mosaic image contains 239 stars detected at H, 121 stars detected
at K, and 119 stars that are detected in both filters. The detected star
closest to the embedded young stellar object is at $14''$ radius. The
reddest detected star is at $17''$ radius and has an $(\hk)$ color of 4.0,
corresponding to over 60 magnitudes of equivalent visual extinction.

The major features of Figure~\ref{fig:nicmos} are:
(1) a strong gradient in the $(\hk)$ colors as one approaches the
protostar;
(2) fewer stars are detected close to the center, despite better
sensitivity;
(3) reddening of the stars turns off into background noise beyond the
$125''$ (0.15~pc) outer radius of dense gas determined from molecular
line data (Zhou et al.\ 1990);
(4) there is a marked asymmetry in the stellar colors associated with
the bipolar outflow-- stars viewed within $\sim45^{\circ}$
of the outflow axis have systematically lower colors;
(5) the dispersion of the $(\hk)$ colors at a given radius and in the
quadrants away from the outflow is $\sim 20\% \times \ehk$ greater than
that of the background population.

\subsection{Theoretical Models of the B335 Density Distribution}
We describe several theoretical models for the density distribution of
B335 and evaluate the success of these models in light of the
extinction data. The models considered are not meant to comprise an
exhaustive list.  They include the inside-out collapse models
previously proposed for B335, Bonnor-Ebert spheres, and simple power
law descriptions. We also discuss briefly the effects of the bipolar
outflow and turbulent motions in B335.

\subsubsection{Inside-Out Collapse Models}
There is a substantial literature of theoretical work by Frank Shu and
colleagues describing the density structure of collapsing dense cloud
cores, starting from the initial condition of a singular isothermal
sphere. Taking parameters appropriate for the B335 region, we briefly
describe the pre-collapse state, the spherical inside-out collapse,
and the modifications to spherical collapse expected for slow initial
rotation or weak magnetic fields.

{\em Pre-collapse Configuration}:
The initial state is that of a singular isothermal sphere; the density
falls off as $\rho \propto r^{-2}$, where $r$ is the radius, with the
normalization of the density determined by the effective sound speed.
For B335, Zhou et al.\ (1990) derive a value for the effective sound
speed of $a=0.23$~km~s$^{-1}$ from molecular line data, which
corresponds to a kinetic temperature of 13 K. The static initial
condition is therefore given by: 
\begin{eqnarray}
\rho_{\mathrm{static}}(r) & = & \frac{a^2}{2 \pi G r^2}  \\
& = & 1.33 \times 10^{-20} \; \mathrm{g} \, \mathrm{cm}^{-3} \; 
      \left(\frac{a}{0.23 \, \mathrm{km} \, \mathrm{s}^{-1}} \right)^2 
      \left(\frac{r}{0.1 \, \mathrm{pc}} \right)^{-2} \\ 
n_{H_2}(r) & = & 2.80 \times 10^{3} \; \mathrm{cm}^{-3} 
\; \left(\frac{a}{0.23 \, \mathrm{km} \, \mathrm{s}^{-1}} \right)^2 
\left(\frac{r}{0.1 \, \mathrm{pc}} \right)^{-2} \ , 
\end{eqnarray}
where the conversion to a molecular hydrogen number density assumes
0.7 for the the mass fraction in hydrogen. 

{\em Spherically Symmetric Collapse}:
The simplest scenario is the spherically symmetric collapse (Shu
1977). In this model, a spherical wave of collapse propagates outwards
from the center at the effective sound speed. The radial distance that
the wave has traveled, sometimes called the infall radius, is the only
additional parameter that defines the density distribution. Inside the
infall radius, conditions approach free fall, with the density taking
the asymptotic form $\rho \propto r^{-3/2}$. By reproducing essential
features of several molecular line profiles with this model, Zhou et
al.\ (1993) claimed evidence for inside-out collapse in B335. Choi et
al.\ (1995) made more detailed radiative transfer models to match the
Zhou et al.\ (1993) observations of H${}_2$CO and CS molecules,
determining the infall radius to be $0.030$~pc ($25''$). Note that
this derived infall radius falls entirely within the deep central
mosaic of NICMOS coverage.

{\em Rotation}:
The effect of slow rotation on the collapse has been solved using a
perturbative approach from the spherically symmetric case (Terebey,
Shu \& Cassen 1984). The result is a distortion to the shape of the
collapse wave of order $4 \times 10^{-4} \, \tau^2$, where $\tau$ is
the number of times that the cloud has rotated since collapse
initiated (this is also equivalent to the ratio of the infall radius
to the centrifugal radius). For B335, Frerking et al.\ (1987) found an
angular velocity of $1.4 \times 10^{-14}$~s${}^{-1}$ through
C${}^{18}$O observations. This indicates a value of $\tau^2 \sim 3
\times 10^{-3}$, and a distortion in the wave by only $\sim 10^{-6}$
--- an entirely negligible effect. The density distribution is more
strongly distorted, with the effect largest near the center of
collapse. For radii greater than a tenth of the infall radius (i.e.\
$r \ge 2\, \farcs5$) --- already well within the innermost radius
where stars are detected --- the correction is always less than of
order $\tau^2$, which is also completely negligible. Such slow
rotation also has an unimportant effect on the model line profiles
towards the center of B335 (Zhou 1995). We therefore do not include
rotation in our models.

{\em Magnetic Fields}:
The effect of a weak magnetic field on the inside-out collapse of a
cloud has been studied by Galli \& Shu (1993) and Li \& Shu (1997).
The collapse wavefront and the density distribution are again
distorted, in a similar way to the rotational model. In the magnetic
case, the parameter $\tau$ is given by the ratio of the infall radius
to the magnetic radius: 
\begin{eqnarray}
R_m & = & \frac{2\, a^2}{B_0 \, \sqrt{G}} \\
    & = & 0.044 \, \mathrm{pc} \, \left(\frac{a}{0.23 \, \mathrm{km} 
\, \mathrm{s}^{-1}} \right)^2 \, \left(\frac{B_0}{30 \, \mu 
\mathrm{G}} \right)^{-1}
\end{eqnarray} 
Galli \& Shu (1993) showed that the field effects only a small
perturbation to the spherically symmetric solution outside a radius of
$r \ge 0.2 \, \tau^{4/3} \, R_{\mathrm{inf}}$.
Figure~\ref{fig:nicmos} shows that variation in color between the two
spatial populations of stars can be as large as a factor of two, well
beyond the infall radius of $25''$ derived by Choi et al.\ (1995).
Since this certainly cannot be construed as a ``small perturbation''
at these radii, we can place a lower limit on the magnetic field that
would be needed to explain the asymmetry in the context of this model.
Taking $r/R_{\mathrm{inf}} \simeq 1$, we find that:
\begin{equation}
B_0 \gtrsim 150 \, \left( \frac{a}{0.23\,\mathrm{km} \, 
\mathrm{s}^{-1}} \right)^2 \, \left(\frac{R_{\mathrm{inf}}}{25''}
\right)^{-1} \, \mu \mathrm{G}
\end{equation}
Such a minimum magnetic field strength is implausibly large. For
example, Crutcher \& Troland (2000) measured a line-of-sight magnetic
field of B${}\sim 11$~$\mu$G in the starless protostellar core L1544
using the Zeeman effect in lines of OH, and this is much higher than
the typical upper limits found for other dark clouds (Crutcher et al.\
1993). In this context, it therefore seems very unlikely that a
magnetically modified inside-out collapse could be responsible for the
large asymmetry observed in the B335 extinction data.

\subsubsection{Bonnor-Ebert Models}
Bonnor-Ebert models are pressure confined isothermal spheres, for
which the solution remains physical at the origin (Ebert 1955, Bonnor
1956). In common with the singular isothermal sphere, the initial
condition for inside-out collapse, they are solutions of a modified
Lane-Emden equation (Chandrasekhar 1967):
\begin{equation}
\frac{1}{\xi^2} \frac{d}{d \xi} \left( \xi^2 \frac{d \psi}{d \xi}
\right) = \exp{(-\psi)} \ ,
\end{equation}
where $\xi=(r/R_0)=(r/a) \sqrt{4 \pi G \rho_c}$ is the dimensionless
radius, and $\psi( \xi)=- \ln{ (\rho/ \rho_c )}$ is a logarithmic
density contrast, with $\rho_c$ the (finite) central density. Unlike
the singular solution, the Bonnor-Ebert solutions do not diverge at
the origin. Instead, the boundary conditions are that the function
$\psi$ and its first derivative are zero at the origin.

The above equation can be solved by division into two first order
equations which can then be tackled simultaneously using numerical
techniques (in our case a 4th order Runge-Cutta method). There is a
family of solutions characterized by a single parameter --- the
dimensionless outer radius of the sphere, $\xi_{\mathrm{max}}$. For a
given sound speed and a particular choice of the shape of the density
curve (i.e.\ $\xi_{\mathrm{max}}$), there is one additional degree of
freedom, the physical scale of the model. This additional degree of
freedom may be removed by implementing an additional constraint, for
example by fixing the outer radius, the total mass, or the external
pressure.

Configurations with dimensionless outer radius $\xi_{\mathrm{max}} >
6.5$ are unstable to gravitational collapse (Bonnor 1956). The
gravitational collapse of Bonnor-Ebert spheres has been studied
numerically by Foster \& Chevalier (1993). They find that the flow
asymptotically approaches the Larson-Penston solution at the origin at
the time of and prior to the formation of a central core, but
emphasize that these large early accretion rates last for a very short
time. If the cloud is initially very centrally condensed (i.e.\
$\xi_{\mathrm{max}} \gg 6.5$) the later stages of infall closely
resemble Shu's inside-out collapse of a singular isothermal sphere.

Recently, Alves, Lada \& Lada (2001) used the color excess method to
study B68, a starless core, with data from the ESO VLT.  The B68
system is much less centrally condensed than B335, and has much lower
column density at the center (only about 30 magnitudes of equivalent
visual extinction). The Alves et al.\ (2001) extinction map of B68
suggests that the density structure is well described by a
Bonnor-Ebert sphere with dimensionless outer radius slightly in excess
of critical: $\xi_{max}=6.9 \pm 0.2$. Although the B335 core is at a
much later evolutionary stage than B68, with an embedded protostar, we
consider the Bonnor-Ebert sphere since it describes B68 so
successfully.

\subsubsection{Bipolar Outflow and Turbulence}
The outflows from young stellar objects can have a substantial impact
on their surroundings. A bipolar outflow in B335 was discovered in CO
by Frerking \& Langer (1982). Later studies in various tracers by
Cabrit, Goldsmith \& Snell (1988), Hirano et al.\ (1988, 1992),
Chandler \& Sargent (1993), and Wilner et al.\ (2000) show that on
arcsecond to arcminute scales, the outflow is well collimated with a
semi-opening angle of about $25\pm5^{\circ}$, and with the axis lying very
nearly in the plane of the sky. The outflow orientation is close to
east-west, with position angle roughly $100^{\circ}$ in the coordinate
system of the NICMOS images. Inside the region affected by the
outflow, the density structure is modified as the core material is
swept out by the flow, but the exact form of the density field
associated with the outflow is not known.

The bipolar outflow may be an important driver of turbulent motions.
Zhou et al.\ (1990) determined a turbulent component to the effective
sound speed in B335 of $0.08$~km~s$^{-1}$. Coupling of the velocity
and density fields via the continuity equation suggests that the
perturbations to the velocity field should induce density contrast
perturbations, i.e.\ $\delta \simeq [(1/3) < \! v/a \! > ^2]^{1/2}
\simeq 0.2$. The resulting perturbations to the column densities (and
extinctions) will depend on the size scale of the velocity and density
field perturbations. The perturbations to the column density will be
largest when the size scale is a factor of a few smaller than the
dense core. The extinction map is consistent with this possibility
since the observed colors of neighboring stars are generally very
similar, yet there exists a $\sim20$\% dispersion for stars that are
not neighbors but are located at similar radial distances from the
protostar (see Figure~\ref{fig:nicmos}). The effect of the column
density perturbations are inevitably reduced by integration along the
line-of-sight, but the observed residual turbulence most likely
contributes significantly to the dispersion in the observed colors.

\subsection{Fitting Model Parameters and Evaluating Fit Quality}
We evaluate the goodness of fit for these various model density
distributions by calculating a reduced $\chi^{2}$, defined as:
\begin{equation}
\chi^2_\nu=\frac{1}{N-m} \sum{ \left( \frac{\ehk_{i}^{\, \mathrm{NIC}} -
\ehk_{i}^{\, \mathrm{model}}}{\sigma_{i}^{\, \mathrm{NIC}}} \right)^2}
\ ,
\end{equation}
where $m$ is the number of free parameters in the model being fitted,
and the sum extends over a particular subset of $N$ stars taken from
the total number detected in both NICMOS filters. The values of the
observed color excess $\ehk_{i}^{\, \mathrm{NIC}}$ and the uncertainty
$\sigma_{i}^{\, \mathrm{NIC}}$ are calculated using the properties of
the unreddened stellar background population (see
Section~\ref{sec:background}, and Figure~\ref{fig:backcolor}),
assuming each star to have an intrinsic color of $(\overline{\hk})_{\,
\mathrm{BG}} = 0.13$, with an uncertainty of $\sigma_{\, \mathrm{BG}}
= 0.16$:
\begin{eqnarray}
\ehk_{i}^{\, \mathrm{NIC}} & = & (\hk)_{i}^{\, \mathrm{NIC}} - 
(\overline{\hk})_{\, \mathrm{BG}} \nonumber \\ 
& = & (\hk)_{i}^{\, \mathrm{NIC}} - 0.13 \ ,\\[3 mm] 
\sigma_{i}^{\, \mathrm{NIC}} & = & \sqrt{\sigma_i^2 + \sigma_{\, 
\mathrm{BG}}^2} \nonumber \\ 
& = & \sqrt{\sigma_i^2 + 0.16^2} \ , 
\end{eqnarray}
where $\sigma_i$ is the uncertainty in the observed $(\hk)$ color
of a given star. This procedure implicitly assumes that the
photometric errors ($\sigma_i$) and the intrinsic $(\hk)$ colors
both have Gaussian distributions. Though the latter condition is
not strictly true, Monte Carlo trials show that the $\chi^2_\nu$
values calculated with the above recipe agree very closely with those
produced when the intrinsic color of each star is instead chosen
randomly from the Gaussian distribution with the same mean and
standard deviation as the sample of background colors. We conclude
that the results of the fitting are not sensitive to this
approximation. 

\begin{figure}[htb]
\figurenum{8} 
\setlength{\unitlength}{1in} 
\begin{picture}(6,5)
\put(-1.0,-3.0){\includegraphics{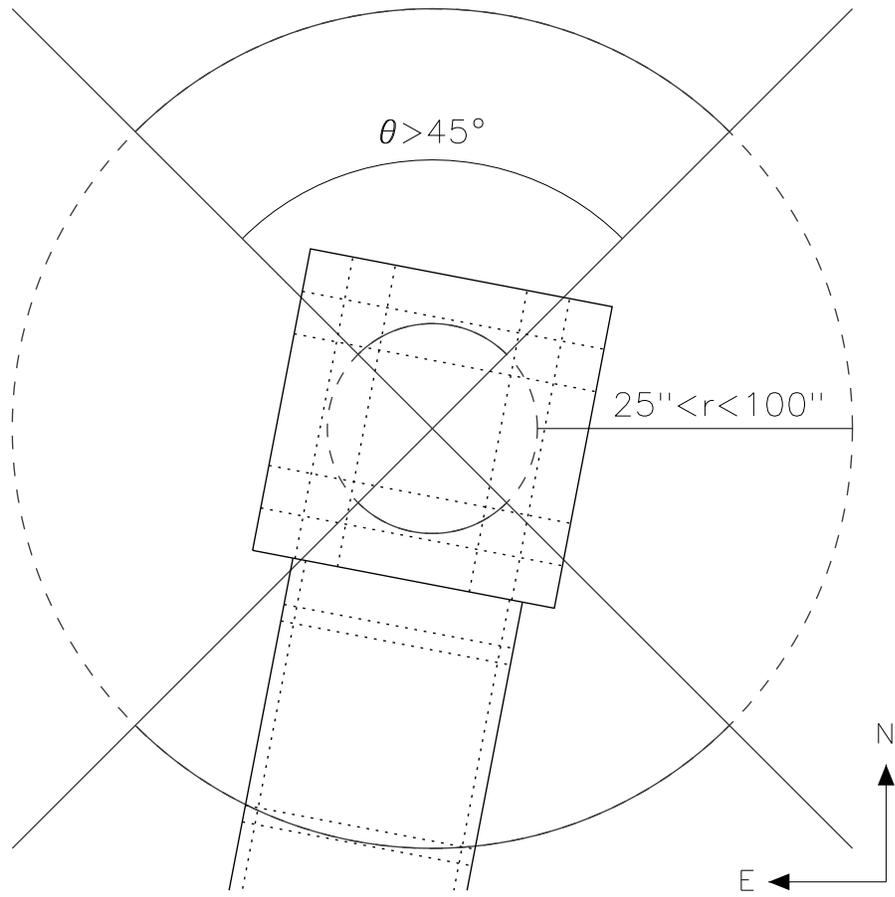}}
\end{picture}
\caption{Schematic diagram showing the scale and coverage of the 
NICMOS observations relative to important scales of B335. The region
beyond $45^{\circ}$ from the outflow axis is marked.
\label{fig:colschem}}
\end{figure}

Since the models are in general non-linear in the fitting
parameters, we analyze the uncertainty in the best-fit model
parameters using a Monte Carlo technique known as the {\em
Bootstrap} method (Press et al.\ 1992). To describe the method, let
us represent the dataset that is being used in the fitting by $S$.
The first step is to construct a new dataset of equal size
$S^{\prime}$, by randomly selecting $N$ times from the original
dataset.  This new sample is then analyzed in the same way as the
original dataset, and the fitted model parameters recorded. This
process is then repeated $n$ times, where $n$ is sufficiently large
that the resulting distribution of best-fit model parameters is
insensitive to its exact value (for our models, typically $n=200$).
The standard deviation of this distribution provides an estimate for
the uncertainty of the parameters that best fit the original
dataset. Table~\ref{tab:fits} summarizes the results from the various
$\chi^2$ analyses and these results are discussed below.
Figure~\ref{fig:colschem} shows a schematic diagram of the spatial
coverage of the NICMOS observations relative to the important size
scales in B335.

\subsection{Excluding the Outflow (Fits A, B \& C)}
The asymmetry in the observed colors appears to be associated with
the bipolar outflow. Since we do not know from theory the effect of
the outflow on the material distribution, we initially restrict
attention to the quadrants that are located away from the flow,
$\theta > 45^{\circ}$, where $\theta$ is the angle subtended between
the vector to a star and the axis of the outflow. To avoid noise
near the boundary of the globule, we also restrict the fitting
region to $r<100''$. In Fit A of Table~\ref{tab:fits}, we use a
power law model for the density distribution. In Fit B we use the
Shu inside-out collapse model, and in Fit C we use a Bonnor-Ebert
sphere.

The best fitting single power law model (Fit A) to the dense core
region unaffected by the outflow has an index of $p=1.91\pm 0.07$.
This value is well matched to the $r^{-2}$ envelope of the isothermal
sphere that represents the initial condition of the inside-out
theory. The inside-out collapse solution (Fit B), which is similar to
a broken power law with fixed indices, provides a somewhat better
description than can a single power law model with arbitrary index.
Remarkably, the fitted infall radius of
$R_{\mathrm{inf}}=28 \pm 3 \, ''$ ($0.034\pm0.004$~pc) agrees within
uncertainties with those calculated by Zhou et al.\ (1993) and Choi
et al.\ (1995) by modeling molecular line profiles. That the minimum
$\chi^2_\nu$ value for these fits are near 3 instead of unity
reflects the $\sim 20\% \times \ehk$ greater dispersion in the
observed colors than in the intrinsic colors of the background stars.

\begin{figure}[htb]
\figurenum{9}
\setlength{\unitlength}{1in}
\begin{picture}(6,5)
\put(-2,-4.7){\includegraphics{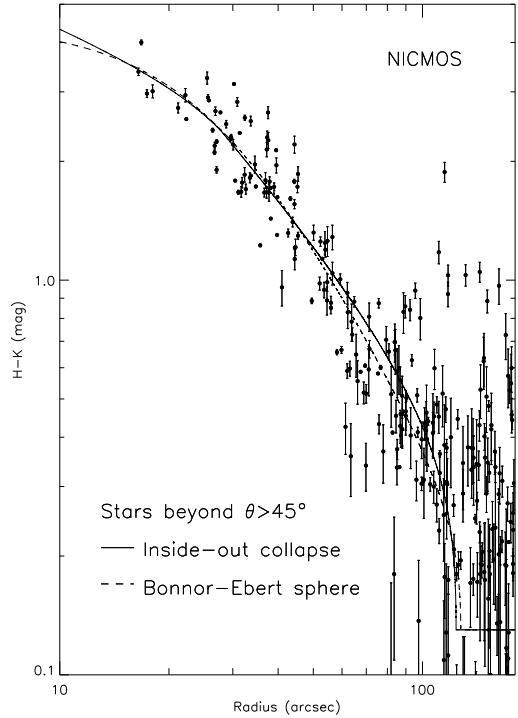}}
\end{picture}
\caption{$(\hk)$ vs.\ radius for the NICMOS stars beyond $45^{\circ}$
from the outflow axis. Overplotted are predicted curves for the best
fitting infall model (Fit B) and the unstable Bonnor-Ebert sphere (Fit
C), assuming a constant background color of 0.13. The two curves are
indistinguishable over the range in radius where there are stars to fit.
\label{fig:fitcomp}}
\end{figure}

A Bonnor-Ebert sphere provides an equally good description of the 
density structure in B335 (Fit C). Figure~\ref{fig:fitcomp} shows a
comparison of the $(\hk)$ radial profiles predicted by the
Bonnor-Ebert sphere in Fit C, and the inside-out collapse model in 
Fit B. The plot shows that the two models are practically
indistinguishable over the range in radius where there are stars to
fit ($14''$--$100''$). At large radii, the Bonnor-Ebert
solution has slope close to a power law index of 2.0, and at smaller
radii, the slope flattens. In effect, the best fit Bonnor-Ebert
sphere is one that mimics the inside-out collapse solution over the
range in radius for which there is data. The models diverge only at
such small radii that extinction data of sufficient depth to
discriminate between them is currently impossible to obtain. One 
discriminant may be sensitive, high resolution observations of dust
emission, which is strongest at these small radii
where extinction data is lacking. Regardless, the value of
$\xi_{max}=12.5\pm2.6$ for the best fit Bonnor-Ebert sphere is well 
in excess of the critical value of 6.5, and it is therefore an unstable
configuration that should undergo gravitational collapse similar
to Shu's inside-out solution (Foster \& Chevalier 1993). 

\subsection{Modeling the Outflow}
\label{sec:outflow}
\begin{figure}[htb]
\figurenum{10}
\setlength{\unitlength}{1in}
\begin{picture}(6,2.2)
\put(-1.2,-7.77){\includegraphics{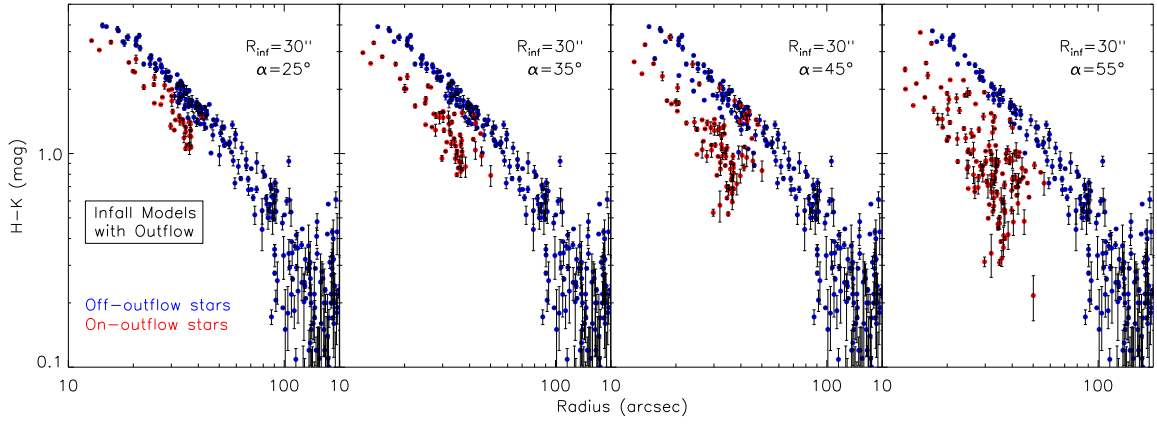}}
\end{picture}
\caption{$(\hk)$ vs.\ radius for simulations of an infalling model with 
outflow cones of various outflow semi-opening angles, $\alpha$.
\label{fig:outflow}} 
\end{figure}

The simplest model we imagine to simulate the effects of the bipolar
outflow on B335 is to assume that the outflow has completely cleared
out the dense material in a bipolar cone of constant opening angle.
Figure~\ref{fig:outflow} shows the radial behavior of the $(\hk)$
colors from simulated observations we have made of an infalling model
for four different values of the model outflow semi-opening angle
$\alpha$. (The appendix describes how the simulations are constructed.) 
Stars viewed through the model outflow cone are marked in red, and
those not viewed through the outflow cone are marked in blue.
The density distribution for $\theta>\alpha$ in each of the simulations
is the infalling model that best fits the $\ehk$ data beyond
$45^{\circ}$ from the outflow axis (Fit B).
As the model opening angle increases, the stars begin to separate into 
two distinct populations. Two effects are evident as the angle increases:
(1) the population of detected stars with reduced colors becomes larger, 
and (2) the separation in the colors of the two populations grows larger.
Manifest in the simulation with the largest opening angle
($55^\circ$) is that the first effect is not only caused by the outflow
cone occupying a larger fraction of the sky, but is also due to there
being less column density for lines-of-sight passing through the cone, 
which allows a greater number of the background stars to be detected. 
Comparison of the radial behavior of the $(\hk)$ colors in
these simulations with the observations in Figure~\ref{fig:nicmos}
shows that these simple models successfully reproduce the major features. 
The extent of the spread between the colors of the two populations of stars 
in the NICMOS data seems to be reproduced best by the simulations with 
semi-opening angles of $35^{\circ}$ and $45^{\circ}$. 
These angles also roughly reproduce the observed number of stars with 
significantly reduced colors.

\subsection{Including the Outflow (Fits D \& E)}
\label{sec:sims}
The simple hollow cone model of the bipolar outflow allows the stars 
viewed closer to the outflow axis to be included in the $\chi^2$
analyses. The fact that the predicted $(\hk)$ colors are a sensitive 
function of the model outflow opening angle means that we are able to fit 
for the opening angle of the flow.

\begin{figure}[htb]
\figurenum{11}
\setlength{\unitlength}{1in}
\begin{picture}(6,4.)
\put(-0.,-2.3){\includegraphics{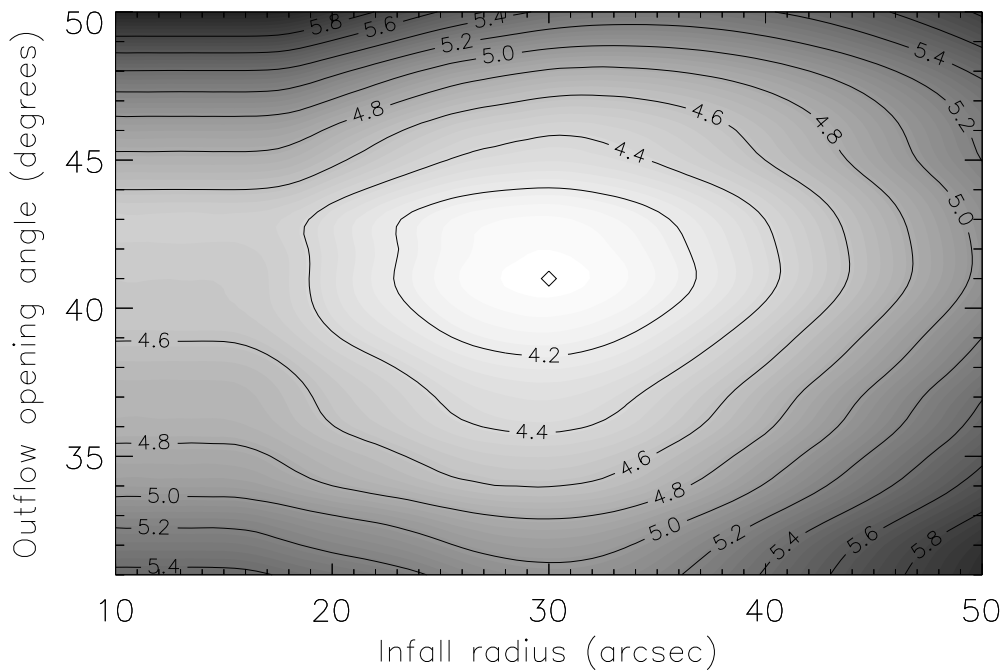}}
\end{picture}
\caption{Contours of the surface of $\chi^2_\nu$ for models with
both infall and outflow, where the outflow has been assumed to clear
out a cone of opening angle ($\alpha$). The minimum in the reduced
$\chi^2$ is 4.02 and corresponds to fitted parameters of
$R_{\mathrm{inf}}=30 \pm 3''$ ($0.034 \pm 0.04 \, \mathrm{pc}$ at a
distance of $250\, \mathrm{pc}$), and $\alpha=41 \pm 4^{\circ}$. The
position of the fitted paramters is marked with a diamond. The best
fitting model is labelled Fit D in Table~\ref{tab:fits}.
\label{fig:chisurf}}
\end{figure}

In both Fits D \& E, we fit for the entire region within $r<100''$,
with a model which has both the infall radius and outflow opening
angle as free parameters, and with the color excess normalization
for each model chosen to minimize the $\chi^2_\nu$ in the region beyond
$45^{\circ}$ from the outflow axis (as in Fit B). The only difference
between Fit D and Fit E is that Fit E includes an extra 
$20\% \times \ehk$ dispersion term in $\sigma_{i}^{\mathrm{NIC}}$ to
account for additional structure in the cloud, perhaps from the
residual turbulent fluctuations. Table~\ref{tab:fits} lists the
best-fit parameters for Fit D and Fit E, and they are nearly
identical. In both Fits D \& E, $\chi^2_\nu$ is essentially
an independent function of the infall radius and the outflow opening
angle. Figure~\ref{fig:chisurf} shows a contour plot of the surface of
$\chi^2_\nu$ for Fit D. The minimized value of $\chi^2_\nu$ is 4.02,
with best fit values of the infall radius and outflow opening angle of 
$R_{\mathrm{inf}}=30 \pm 3''$ ($0.036\pm0.004 \, \mathrm{pc}$) and 
$\alpha=41 \pm 4^{\circ}$, respectively. For Fit E, the contour plots
and fitted parameters are essentially identical--- infall radius 
$R_{\mathrm{inf}}=26 \pm 3''$ ($0.031\pm0.004 \, \mathrm{pc}$) 
and outflow opening angle $\alpha=41 \pm 2^{\circ}$--- but the minimized
$\chi^2_\nu$ is $\simeq 1$ due to the extra dispersion.
That the two fits agree so closely, particularly with regard
to the infall radius, is interesting. The extra dispersion in Fit E
reduces the weight of the most reddened (and most central) stars,
which inevitably are those that constrain the fitting of the infall radius. 

The composite model with infall and bipolar outflow provides a
consistent description of the density structure of B335. When the
regions affected by the outflow are excluded from consideration, the
inside-out collapse solution provided a good description of the data.
Modeling the outflow as a hollowed-out bipolar cone allows inclusion
of the stars from the outflow region in the $\chi^2$ analysis. Despite
the obvious crudeness of this model for the outflow, the fits to
the entire dataset (Fits D \& E) constrain both the infall radius 
and outflow opening angle, and the fitted infall radii are entirely
consistent with our earlier determination (Fit B) and with the values
determined by molecular line studies. The fitted opening angle of the
outflow is somewhat larger than indicated in studies of high velocity
CO emission. Examination of the CO velocity structure 
(e.g.\ Hirano et al.\ 1988, 1992) 
suggests that the flow is more collimated at higher velocities, and 
that the opening angles dervied from CO must be lower limits since the 
low velocity outflow becomes confused with the ambient globule emission 
(and is resolved out by interferometers). 
This effect may explain why the opening angle appears wider in extinction 
than in CO emission.
Of course, the one parameter model fitted here simply may not provide 
an adequate description of the outflow geometry at this level of detail.

\begin{figure}[h]
\figurenum{12}
\setlength{\unitlength}{1in}
\begin{picture}(6,8.)
\put(-0.6,-2.1){\includegraphics{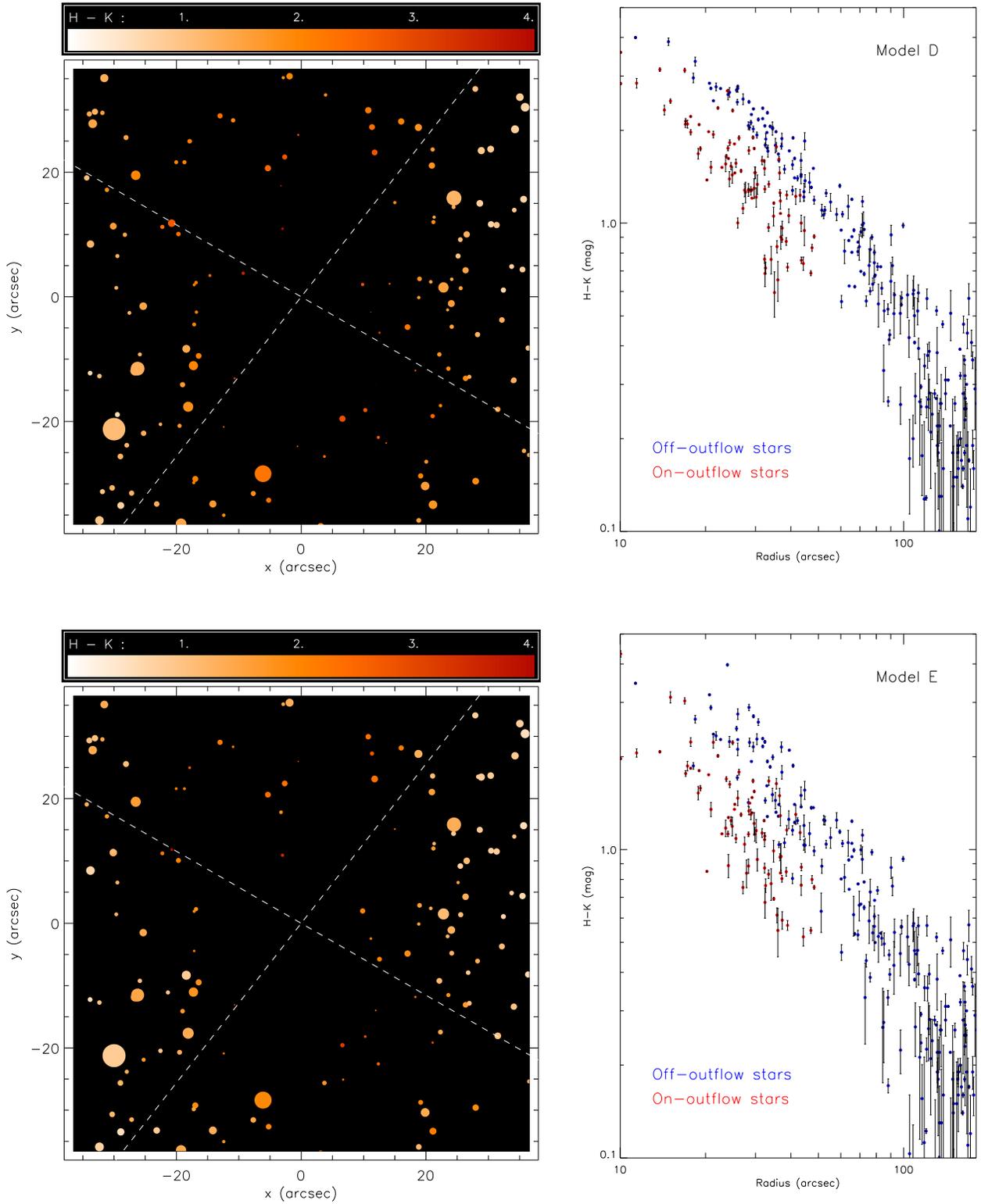}}
\end{picture}
\caption{A simulation of the models of Fit D (Top) and Fit E (Bottom).
For each model: (Left) Pseudo-image of the simulated NICMOS mosaic image; 
(Right) $(\hk)$ color vs.\ radius out to $180''$ from the center.
\label{fig:models}}
\end{figure}

Figure~\ref{fig:models} presents
a simulation of the models for Fit D and Fit E to afford another way
to examine the success of these models, as well as the differences 
between them (the appendix describes the detailed recipe for constructing 
these simulations). The model for Fit D 
has an infall radius of $30''$ and an outflow opening angle of 
$41^{\circ}$ (the normalization of the color excess is matched to the data 
by the $\chi^2$ analysis). This model reproduces the gross structure of 
the NICMOS data very successfully. However, the radial plot shows that 
the dispersion in the $(\hk)$ colors at a given radius in the simulation
is clearly significantly less than in the observations. The model for Fit E 
is nearly identical to Fit D except that the color excess has been multiplied 
by a Gaussian random variable with mean of unity and standard deviation 20\%. 
The simulated data from the model for Fit E,
with this additional dispersion to mimic the effects of residual
turbulence, compares much more favorably with the observations in
Figure~\ref{fig:nicmos}.

\subsection{The Scaling Factor}
\label{sec:scaling}
\begin{figure}[h]
\figurenum{13}
\setlength{\unitlength}{1in}
\begin{picture}(6,4.0)
\put(-1.2,-6.){\includegraphics{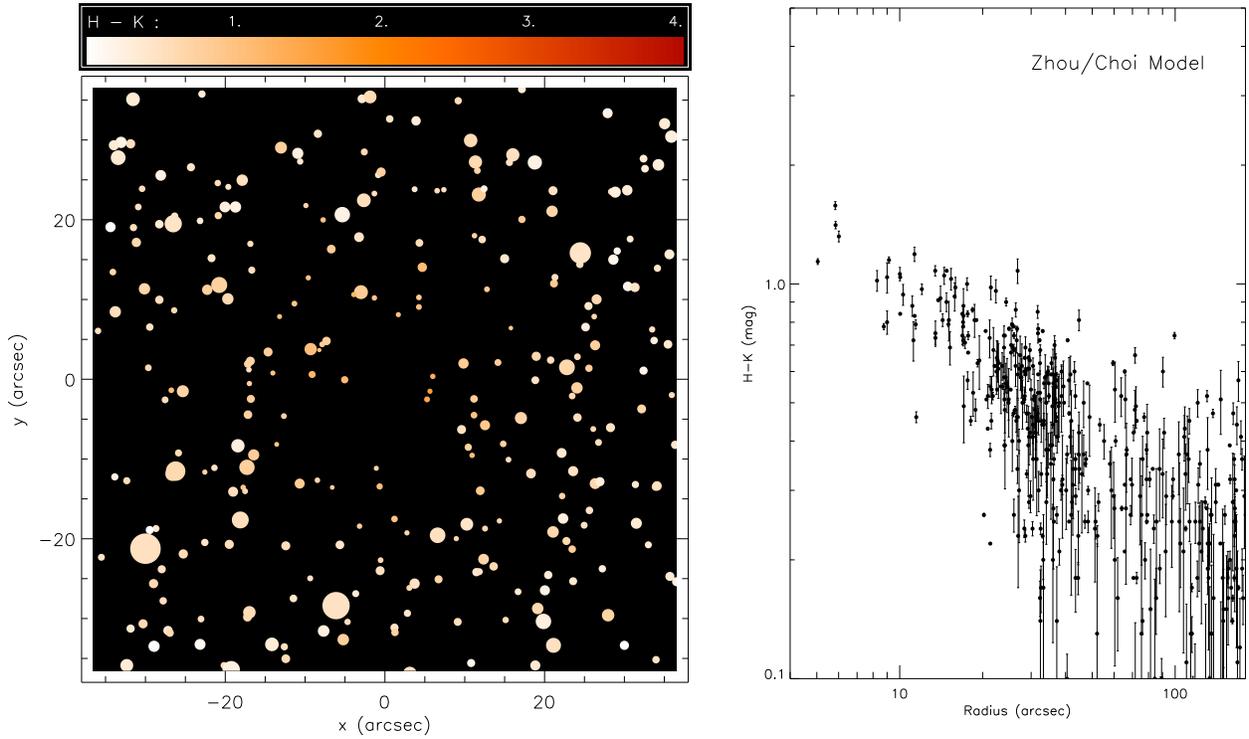}}
\end{picture}
\caption{A simulation of the spherically symmetric inside-out collapse
model with infall radius $25''$ (Choi et al.\ 1995), and normalization
determined by the sound speed of $a=0.23$~km~s${}^{-1}$ (Zhou et al.\
1990) and standard distance of 250~pc (Tomita et al.\ 1979). Left:
Pseudo-image of the simulated NICMOS mosaic image. Right: $(\hk)$ color
vs. radius out to $180''$ from the center.\label{fig:modelb}}
\end{figure}

So far, we have been concerned only with the {\em shape} of the
density profile and have ignored the free scaling factor applied to
the color excess predicted by the various models during the
fitting. For both the inside-out collapse models and the Bonnor-Ebert
spheres, the normalization of the density distribution is entirely
determined by the effective sound speed, measured to be
$0.23$~km~s$^{-1}$ (Zhou et al.\ 1990). This measurement, together
with the distance to the cloud, gas-to-dust ratio, and near-infrared
reddening law, make a unique prediction for the color excess
normalization in each model. As listed in Table~\ref{tab:fits}, a
scaling factor $\mathcal{F} \sim{}$3--5 is required to match the
models with the observed color excess. Figure~\ref{fig:modelb} shows
a simulation of the spherically symmetric inside-out collapse model
with the effective sound speed of $0.23$~km~s${}^{-1}$, the
standard distance of 250~pc (Tomita et al.\ 1979), and the infall
radius of $25''$ derived by Choi et al.\ (1995).
The simulation demonstrates the extent to which the
model underestimates the observed reddening and extinction in B335. 
If this model were correct, then the NICMOS observations would have
been sufficiently deep to see right through the center of the cloud,
and the number of stars detected would have been far greater than
observed. This therefore leads to the question: if the inside-out
collapse model or the Bonnor-Ebert sphere model is correct, 
then what explains the need for this large scaling factor between 
the predicted extinction and the measured values?

The first possibility to consider is that the effective sound speed in
B335 has been underestimated. To account for the $\mathcal{F} \sim 
{}$3--5 by a larger kinetic temperature in the initial state seems
implausible --- a temperature of $T\simeq {}$40--60~K would be
needed. Multi-transition observations of NH${}_3$ show that the bulk
of the dense gas in B335 has a kinetic temperature of 10--12 K (Menten
et al.\ 1984), consistent with a balance between heating and cooling
processes. To account for the factor with non-thermal motions also
seems implausible, given the measured widths of molecular lines. A
non-thermal contribution of $v_{\mathrm{turb}}\simeq 
{}$0.35--0.45~km~s$^{-1}$ would be required, which is clearly at odds
with a multitude of molecular line data. For example, Zhou et al.\
(1990) used observations of H${}_2$CO to determine a small turbulent
contribution of $v_{\mathrm{turb}}=0.085$~km~s$^{-1}$ to the thermal
sound speed. Similar results from a variety of different molecules are
documented by Menten et al.\ (1984) and Frerking, Langer \& Wilson
(1987). One might envision a form of non-thermal support that is not
reflected in the observed molecular line-widths, perhaps because of a
special viewing geometry. If magnetic support were primarily from a
static field and not waves, and this field were oriented so that the
motions were not observable along the line-of-sight, then it would be
effectively undetectable. Although such a special situation seems
contrived and unlikely, the orientation of B335 is special in that 
the outflow lies close to the plane of the sky, which may make such 
a configuration more plausible.

Another possibility that affects the overall scaling lies in the outer
boundary, i.e. where the cloud ends. Zhou et al.\ (1990) calculated
the mass of B335 using an outer radius of 0.15~pc based on their radio
map and the 250~pc distance estimate. The molecular line profiles of
dense gas tracers are not particularly sensitive to this assumption.
We have adopted this value for the outer radius throughout the
extinction modeling. Indeed, the column densities are also
insensitive to this assumption since the density falls off steeply
with radius. For a star at the infall radius of the best-fitting
inside-out collapse model (Fit B), doubling the outer radius of the
cloud increases the predicted extinction by only 7.5\%, which is an
unimportant amount with respect to the derived scaling factors.

An overestimate of the distance to B335 would also result in predicted
column densities that are too small. This distance uncertainty may be
the dominant source of systematic error in the study. The distance of
250~pc was adopted from optical measurements of stellar reddening
versus distance by Tomita et al.\ (1979), and this distance is
uncertain at the 50\% level. If B335 were closer, then the extinction
would be larger at the same angular radius. To account for the full
scaling factor would require that the distance be $\sim{}$3--5 times
closer, only about 50--80~pc, which seems unreasonable. Also, for a
given level of observed extinction, the inferred mass of the globule
is proportional to the square of its distance (since the angle
subtended by B335 is fixed). For the outer boundary assumption of
Zhou et al.\ (1990), the mass inferred from our extinction data is
\begin{equation} 
M_{\mathrm{B335}} \simeq 14 \,
\left(\frac{d}{250\mathrm{pc}}\right)^2 \:  \mathrm{M}_{\odot} \ .
\end{equation} 
The 14~M$_{\odot}$ for a distance of 250~pc may be large for a small
globule, but reducing the distance by the whole scaling factor lowers
the mass to less than 1~M$_{\odot}$ for the inside-out collapse model,
and a little over 1~M$_{\odot}$ for the Bonnor-Ebert sphere. 
This seems small given that the globule must be forming a star with 
a substantial fraction of a solar mass, and losing a lot of mass
through outflow in the process. Furthermore, in the context of the
inside-out collapse model, a reduction in the distance also results
in a smaller mass for the protostellar core, as well as a shorter time
since the onset of collapse (both are proportional to the distance for
a given sound speed and infall radius). In short, a closer distance for
B335 would ameliorate the discrepancy between the observed and predicted 
extinction, but it is not clear if this could account for the entire 
difference.

A final uncertainty is the conversion from column density to color
excess using the gas-to-dust ratio and reddening law. Significant
variations in the reddening law have not been observed, and they are
unlikely to explain much of the observed discrepancy in the color
excess (although we briefly return to this point in the appendix). By
contrast, the gas-to-dust ratio has been observed to vary in the
Galaxy, with variations of $\sim 2$ not uncommon along different lines
of sight (Bohlin et al.\ 1978). If the gas-to-dust ratio in B335 were
lower than the standard value for the interstellar medium, or the
reddening law steeper, then this would cause us to underestimate the
extinction. Potentially, the full scaling factor could be accounted
for by this uncertainty, or perhaps by some combination of a closer
distance and a modified gas-to-dust ratio. For example, if the true
distance to B335 were actually 100~pc, and the gas-to-dust ratio were
about one half of the standard value (i.e.\ $R=1\times10^{21} \,
\mathrm{cm}^{-2} \, / \mathrm{mag}$), then these two factors would
fully account for the discrepancy between observed extinction and the
predictions of either model, and also give a reasonable mass for the
globule of $M\simeq 2 ~\mathrm{M}_{\odot}$. If molecular hydrogen were
to deplete onto dust grains, this could cause such a decrease in the
gas-to-dust ratio at high density. However, H${}_2$ is not thought 
to undergo significant freeze-out in this way (Sandford \&
Allamandola 1993), and a decrease in the gas-to-dust ratio is at 
odds with evidence that suggests an increase in high extinction regions
(e.g.\ Bohlin et al.\ 1978, de Geus \& Burton 1991, and Ciolek \&
Mouschovias 1996).

\subsection{Comparison with Dust Emission Studies}
The density distribution in protostellar cores may also be studied 
through dust emission. If the emission is observed at long enough
wavelength that the Rayleigh-Jeans approximation applies, then, in the
optically thin regime, the intensity at impact parameter $b$ is
proportional to the integral along the line-of-sight of the product of
the density and the temperature. Assuming that the intensity, temperature
and density obey power law distributions in radius, then the power law index 
of the density distribution $\rho \propto r^{-p}$ is simply given by 
$p=m+1-q$, where $m$ is the observed power law index of the intensity profile, 
and $q$ the power law index of the temperature profile. 
The mass may also be derived from the radial intensity distribution, 
given an assumed mass opacity. Unfortunately, the mass opacity of dust at 
millimeter and submillimeter wavelengths is uncertain, by a factor of five 
or more (Ossenkopf \& Henning 1994).

For B335, dust emission at submillimeter and millimeter wavelengths
has been studied recently by Shirley et al.\ (2000) and Motte \&
Andr\'{e} (2001).  Shirley et al.\ (2000) constructed images at
850~$\mu$m and 450~$\mu$m at resolutions of $15''$ and $8''$,
respectively, using SCUBA at the JCMT. At both wavelengths, these
observations are sensitive to emission out to an angular radius of
roughly $52''$.  They adopt a temperature profile with power law index
$q=0.4$ for a region where the envelope is optically thin to the bulk
of the infrared radiation (e.g.\ Emerson 1988; Butner et al.\ 1990).
They fit the circularly averaged maps at each wavelength, extending
from inner cutoffs at angular radii of $24''$ for the 850~$\mu$m
profile and $12''$ for the 450~$\mu$m profile. They derive power law
indices of $p=1.74 \pm 0.4$ and $p=1.65 \pm 0.17$ from the 850~$\mu$m
and 450~$\mu$m emission, respectively. They estimate a mass for B335
of 1.2~M$_{\odot}$ using the total 850~$\mu$m flux in a $120''$ area
and assuming a sphere of constant density.
Motte and Andr\'{e} (2001) constructed a 1.3~mm continuum image of
B335 with $11''$ resolution using the IRAM 30m telescope, detecting
emission out to an angular radius of $120''$. They adopt an isothermal
temperature distribution and calculate a density power law index of
$p=2.2 \pm 0.4$. In this model, the total mass is $2.5\,
\mathrm{M}_{\odot}$. They note an asymmetry in the emission, with the
emission being elongated perpendicular to the outflow axis. As in this
extinction study, they perform fits to the off-outflow regions, and as
a result increase the uncertainty in their fitted density index by 0.2
(included in the quoted uncertainty). 

Taking account the uncertainties, the density power law indices
derived from the millimeter and submillimeter emission agree both with
each other and with the results from dust extinction analysis (Fit A,
$p=1.91\pm0.07$).  However, the extinction method provides a direct
measurement of the distribution that suffers none of the added
uncertainties associated with the temperature profile in the core.

\section{Summary}
\label{sec:conc}
We present a near-infrared extinction study of the protostellar
collapse candidate B335 using deep integrations from HST/NICMOS
supplemented by data from the W.M.\ Keck Observatory. In summary:

\begin{enumerate}
\item We determine transformations between the NICMOS filters F160W
and F222M and the standard H and K bands in the CIT photometric system
for a range of $(\hk)$ color from 0 to 3. At the accuracy of our data,
the derived transformations do not have significant slopes or color
terms.

\item The NICMOS mosaic image of the central $72''$ region of B335
shows a dramatic fall off in the number of stars detected towards the
protostar location, where the extinction increases due to the central
concentration of dense core material.  The NICMOS mosaic image
contains a total of 119 stars detected in both filters, the innermost
of which is at a radius of $14''$. The photometry shows a steep
gradient in the $(\hk)$ colors toward the central protostar.  In
addition, there is a strong asymmetry in the colors superimposed on
the overall gradient that coincides with the bipolar outflow.

\item We compare a series of models of dense core stucture to the
extinction data, including the previously proposed models of
inside-out collapse derived from molecular line observations. The
{\em shape} of the $\ehk$ radial profile is well matched by the
inside-out collapse model, in particular the density fall-off as
$r^{-2}$ for the envelope and a flattening within the purported infall
zone. An unstable Bonnor-Ebert sphere with dimensionless outer radius
$\xi_{\mathrm{max}}=12.5\pm2.6$ provides an equally good description
of the density profile, and is indistinguishable from the collapse
model over the range in radius where there are stars to fit. 
The inside-out collapse model is not unique. 
The power law index for the density distribution derived
from extinction is consistent with recent determinations from dust emission
measurements at millimeter and submillimeter wavelengths.

\item The bipolar outflow appears to be responsible for the dominant
asymmetry in the extinction data. Stars whose line of sight fall within
$45^{\circ}$ of the outflow axis show lower extinction
than those further away from the outflow. The effect of the outflow on the
extinction data is reproduced very well by models that consider a
hollowed out bipolar cone of constant opening angle. In the context of
the inside-out collapse models, the observed asymmetry is too large to
be explained by either rotation, or a weak magnetic field.

\item The entire extinction dataset can be well fitted by a model
that includes infall, bipolar outflow, and an additional $20\% \times \ehk$
dispersion that mimics residual turbulent motions. 
The best fit outflow semi-opening angle is $\alpha = 41 \pm 2^{\circ}$ and 
the best fit infall radius is $R_{\mathrm{inf}} = 26 \pm 3''$. The fitted 
value of the infall radius is consistent with those derived from 
molecular line studies of B335, and supports the inside-out collapse 
interpretation of the density structure. The fitted opening angle for the 
bipolar outflow is somewhat larger than has been observed in high velocity 
CO emission, perhaps because the molecular line studies do not recover the 
full outflow width due to confusion with ambient emission at low velocities.

\item The normalization of the observed color excess is a factor of
$\mathcal{F}$${}\sim 5$ higher than the value predicted using the
nominal 250~pc distance to B335, and the standard gas-to-dust ratio
and near-infrared reddening law. The discepancy can conceivably be
explained by a combination of a closer distance and a lower
gas-to-dust ratio.
\end{enumerate}

\acknowledgements
Partial support for this work was provided by NASA through grant
number HST-GO-07843.01-A from the Space Telescope Science Institute,
which is operated by AURA, Inc., under NASA contract NAS5-26555 We
thank Bob Goodrich and Randy Campbell for invaluable assistance with
the observations obtained at the W.M.\ Keck Observatory.  DWAH thanks
Brian McLeod for his help in running the NICREDUCE software, and Alex
Wilson for his assistance with IDL graphics.

\appendix

\section{Simulated Extinction Observations}
This appendix contains the tools and general recipe for constructing
the simulated observations of B335 presented in
Sections~\ref{sec:sims}~\&~\ref{sec:scaling}.

\subsection{The Background Luminosity Function}
We construct a composite luminosity function for the background
population by combining the data from fields away from the extinction
of the B335 dense core. In deriving the luminosity function, the
photometry for each field was cut at a very high completeness level,
where the number of stars per unit magnitude range peaks. (This is a
very conservative estimate of limiting magnitude.) To avoid any issues
concerning how the stars should be binned, the cutoff was determined
by finding where the gradient of the cumulative luminosity function
was steepest.  For the dataset from the fifth NICMOS strip field,
which includes 202 detected stars at H and 85 detected stars at K, the
cut-off magnitudes are 20.7 at H, and 18.7 at K. For the combined
dataset of the four Keck off-fields, which includes 242 detected stars
at H and 246 detected stars at K, the cut-off magnitudes were 18.8 for
both filters.

The empirical H-band luminosity function was calculated in three parts
out to the NICMOS high completeness limit (H${}=20.7$). For the bright
end (H${}<12.85$), only stars from the NICMOS fifth strip field are
included, as stars brighter than this limit were saturated in the Keck
images.  For the mid-range ($12.85<\mathrm{H}<18.8$), stars from both
datasets contribute to the function. For the faint end
($18.8<\mathrm{H}<20.7$), above the Keck high completeness limit, only
stars in the NICMOS fifth strip field were included. In each part, the
number of stars was divided by the relevant observed area to build up
the cumulative luminosity function.  In the part where the NICMOS and
Keck data overlap ($12.85<\mathrm{H}<18.8$), the two datasets show
good agreement. The K band luminosity function was constructed
following the same procedure, and it is effectively the same as for
the H band, offset by the mean color of ${\hk}=0.13$ magnitudes.

\subsubsection{Luminosity Function Extrapolation}
The central regions of B335 were observed to a deeper limiting
magnitude than the strip or the $10'$ off fields, by about 4
magnitudes (H band) at the mosaic center, and about 3 magnitudes at
the edge (see Figure~\ref{fig:limplot}). Although the stars in these
central regions of B335 are highly extincted, the empirical background
luminosity function must be extrapolated in order to simulate the
deepest observations.  Since the least extincted star in the NICMOS
mosaic has $\hk=0.47\pm 0.08$, corresponding to an extinction of
$A_{\mathrm{H}}=1.3 \pm 0.2$, the luminosity function must be
extrapolated by about 2 magnitudes at H in the simulations.

The form of the H cumulative luminosity function suggests that a
broken power law provides a good description. The function was 
fitted by minimizing the $\chi^2$ of the fit to the bright-end of 
the luminosity function. The resulting function is given by:
\begin{equation}
N(m<H)= \mathrm{5.4 \: stars~} \, / \, (')^2 \times
\begin{cases}
(\mathrm{H} \, / \, 14.17)^{2.95} & \mathrm{for \ H} \le 14.17 \\
(\mathrm{H} \, / \, 14.17)^{9.56} & \mathrm{for \ H} > 14.17
\end{cases}
\end{equation}

\subsection{A Recipe for Constucting Simulations}
Simulated observations of B335 were constructed by superimposing a
model cloud on a field of background stars with the appropriate
characteristics. The column density was calculated at the position of
each background star and converted into two observable quantities: (1)
an extinction at H band, and (2) a color excess $\ehk$. These
conversions are based on the gas-to-dust ratio, and reddening law.  (A
brief discussion on the uncertainties of these quantities is given in
Section~\ref{sec:method}.)

\begin{figure}[htb]
\figurenum{2} 
\setlength{\unitlength}{1in} 
\begin{picture}(6,7.0)
\put(-0.5,-1.6){\includegraphics{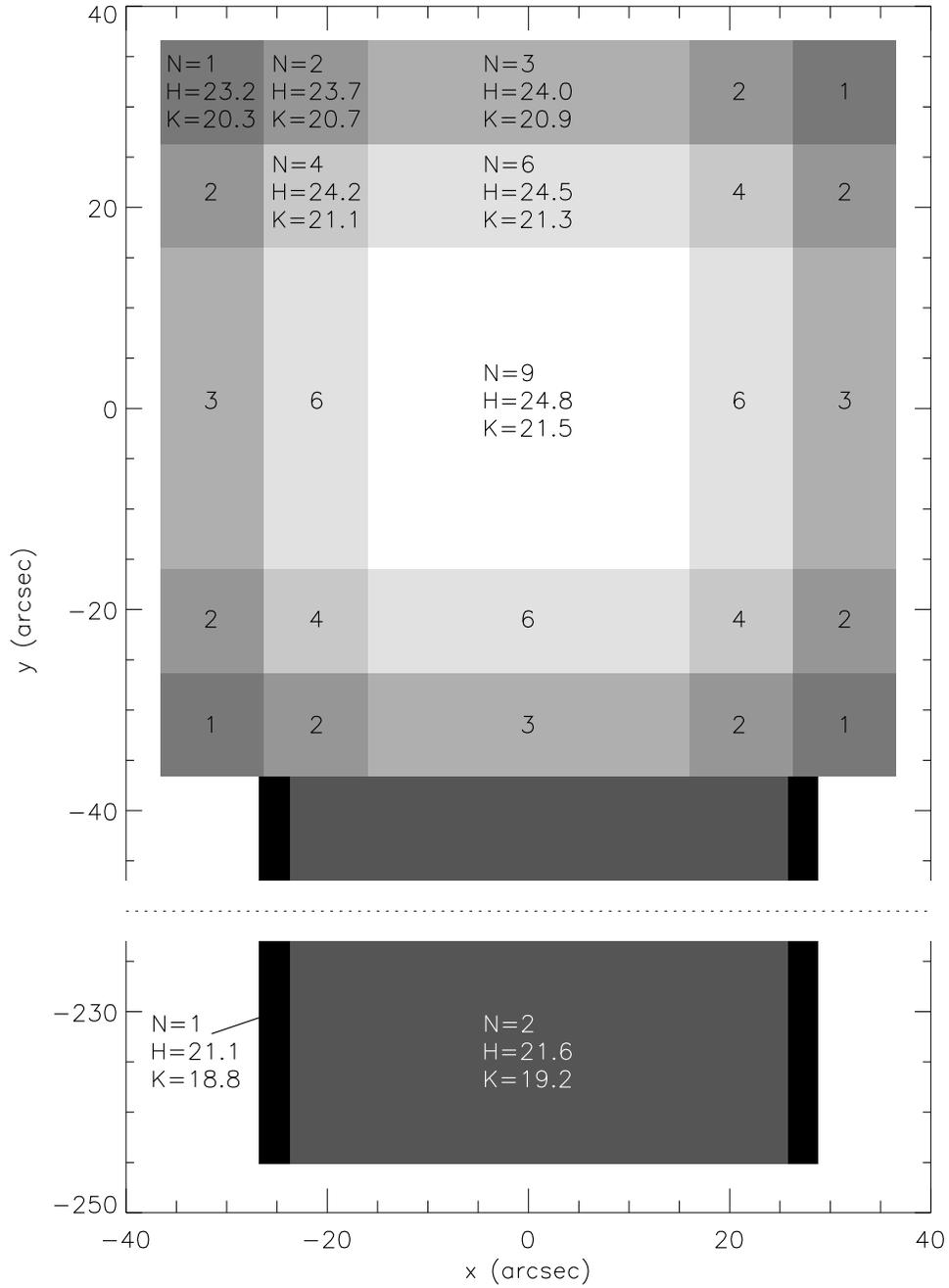}}
\end{picture}
\caption{Diagram showing the spatially dependent limiting magnitudes 
(in the CIT system) for the NICMOS observations. The quantity $N$ indicates 
the number of exposures taken of that region.\label{fig:limplot}}
\end{figure}

To summarize, the recipe for constructing simulated observations has
the following steps:
\begin{enumerate}
\item Generate a background star field by selecting randomly from the
broken-power-law cumulative luminosity function in concert with the
Gaussian color distribution of the background population.  The
broken-power-law luminosity function is extended to H${}=25$, just
beyond the limiting magnitude at the center of the NICMOS mosaic (Note
that model background stars intrinsically fainter than H${}=23$ are
not recovered in the simulated observations of the fitted models from
Table~\ref{tab:fits} --- the function is extended to H${}=25$ simply
to ensure completeness). Positions for each star are assigned randomly
within a region $8'\times2'$, to encompass the area of the NICMOS
observations.
\item Select a model density distribution to describe B335.
\item Calculate the column density for each star in the region. 
\item Convert the column density to an extinction at H band,
$A_{\mathrm{H}}$, and a color excess $\ehk$ using an assumed
gas-to-dust ratio, R, and a reddening law,
$A_{\mathrm{H}}/A_{\mathrm{V}}$. The ``standard'' prescription is
(Bohlin et al.\ 1978, Rieke \& Lebofsky 1985):
\begin{eqnarray}
A_{\mathrm{H}} \: & = \: & 0.88 \: \left(\frac{N(\mathrm{H} + 
\mathrm{H}_2)}{1\! \times \!10^{22} \: \mathrm{cm}^{-2}}\right) 
\left(\frac{2 \! \times \! 10^{21} \: \mathrm{cm}^{-2} 
\mathrm{mag}_V^{-1}}{\mathrm{R}}\right) 
\left(\frac{A_{\mathrm{H}}/A_{\mathrm{V}}}{0.175}\right) \ , \\[3 mm]
\ehk \: & = & A_{\mathrm{H}} \: / \: 2.78 \ .
\end{eqnarray} 
(In the simulations, the reddening law was allowed to vary up to 15--25\% 
steeper than the standard law in order to more exactly reproduce the 
observed starcounts. A fluctuation upward in the star counts at the 
one standard deviation level explains the inferred steeper reddening law.)
\item Modify the magnitudes and colors of each background star
appropriately.
\item Simulate the NICMOS observations by applying the spatially 
dependent detection limits (Figure~\ref{fig:limplot}). 
\end{enumerate}

% The tables and figures begin here....
\begin{deluxetable}{llllll}
\tablenum{1}
\tabletypesize{\scriptsize}
\tablewidth{0pt}
\tablecolumns{6}
\tablecaption{Summary of the $\chi^2$ Analyses Data \label{tab:fits}}
\tablehead{ \colhead{Fit} & \colhead{Model Density Distribution} 
& \colhead{Fit Region} & \colhead{Stars} & \colhead{Fitted Model 
Parameter(s)} & \colhead{$\chi^2_\nu$}}
\rotate
\startdata
A & Scaled Power Law: $\rho \propto r^{-p}$ & $r<100"\, ; \, 
\theta>45^{\circ}$ & 146 & $p=1.91 \pm 0.07$ & 3.37 \\
B & Scaled Infall: $\rho=\mathcal{F}$${} \rho_{\mathrm{Shu}}
(r,R_{\mathrm{inf}})$ & $r<100"\, ; \, \theta>45^{\circ}$ & 146
& $R_{\mathrm{inf}}=28 \pm 3" \, ; \, \mathcal{F}={}$ $5.25 \pm 
0.13$ & 3.10\\ 
C & Scaled Bonnor-Ebert ($R=125''$) & $r<100" \, ; \, 
\theta>45^{\circ}$ & 146 & $\xi_{\mathrm{max}}=12.5 \pm 2.6 \, ; \,
\mathcal{F}={}$ $3.42 \pm 0.10$ & 2.93\\
D & Scaled Infall, Hollow-Cone Outflow & $r<100"$ & 209 &
$R_{\mathrm{inf}}=30 \pm 3" \, ; \, \alpha=41 \pm 4^{\circ} \, ;
\, \mathcal{F}={}$ $5.25 \pm 0.13$ & 4.02\\ 
E & As Fit D with 20\% Dispersion & $r<100"$ & 209 &
$R_{\mathrm{inf}}=26 \pm 3" \, ; \, \alpha=41 \pm 2^{\circ} \, ;
\, \mathcal{F}={}$ $4.77 \pm 0.14$ & 1.04 \\ 
\enddata
\tablecomments{For each Fit, we consider only stars within $100''$ of
the protostar location, to avoid noise-related issues near the edge of
the cloud.  Fits A--C consider only the regions that are well removed
from the outflow, while
Fits D \& E are for the entire dataset within 100''. Fit A is for a
general power law density distribution. The remaining Fits use the
Zhou et al.\ (1990) effective sound speed ($a=0.23$~km~ s${}^{-1}$),
with a free scaling applied to the predicted color excess,
representing the many potential sources of systematic error in the
conversion from color excess to density. Fit B uses the Shu inside-out
collapse model, and Fit C uses a Bonnor-Ebert sphere. Fit D uses a
combination of an infalling model with a model outflow that empties a
cone of constant opening angle. Fit E uses the same model, but
includes an extra dispersion in $\ehk$ to account for extra small
scale structure in the cloud.}
\end{deluxetable}

\clearpage
%Figures

\clearpage 

\clearpage

\clearpage

\clearpage

\clearpage

\clearpage

\clearpage

\clearpage  

\clearpage

\clearpage

\clearpage

\clearpage

\end{document}